\documentclass[aoas]{imsart}

\RequirePackage{amsthm,amsmath,amsfonts,amssymb}
\RequirePackage[authoryear]{natbib}
\RequirePackage[colorlinks,citecolor=blue,urlcolor=blue]{hyperref}
\RequirePackage{graphicx}
\usepackage{bm}
\usepackage{makecell}
\usepackage{threeparttable,booktabs,multirow}
\newcommand{\tabincell}[2]{\begin{tabular}{@{}#1@{}}#2\end{tabular}}
\usepackage[noend, linesnumbered, ruled]{algorithm2e}

\startlocaldefs
\theoremstyle{plain}

\newtheorem{theorem}{Theorem}[section]

\newtheorem{proposition}{Proposition}[section]
\theoremstyle{remark}

\newtheorem{assumption}{Assumption}

\endlocaldefs

\begin{document}
	
	\begin{frontmatter}
		\title{Enterprise Profit Prediction Using Multiple Data Sources with Missing Values through Vertical  Federated Learning}
		%\title{A sample article title with some additional note\thanksref{t1}}
		\runtitle{Enterprise Profit Prediction Using VFEM}
		%\thankstext{T1}{A sample additional note to the title.}
		
		\begin{aug}
			%%%%%%%%%%%%%%%%%%%%%%%%%%%%%%%%%%%%%%%%%%%%%%%
			%% Only one address is permitted per author. %%
			%% Only division, organization and e-mail is %%
			%% included in the address.                  %%
			%% Additional information can be included in %%
			\author[B]{\fnms{Huiyun}~\snm{Tang}\ead[label=e2]{2020103671@ruc.edu.cn}},
			\author[A,B]{\fnms{Feifei}~\snm{Wang}\ead[label=e1]{feifei.wang@ruc.edu.cn}},
			\author[C]{\fnms{Long}~\snm{Feng}\ead[label=e3]{lfeng@hku.hk}}
			\and
			\author[A,B]{\fnms{Yang}~\snm{Li}\ead[label=e4]{yang.li@ruc.edu.cn}}
			%%%%%%%%%%%%%%%%%%%%%%%%%%%%%%%%%%%%%%%%%%%%%%
			%% Addresses                                %%
			%%%%%%%%%%%%%%%%%%%%%%%%%%%%%%%%%%%%%%%%%%%%%%		
			\address[B]{School of Statistics, Renmin University of China, Beijing, China \printead[presep={,\ }]{e2}}
			\address[A]{Center for Applied Statistics, Renmin University of China, Beijing, China \printead[presep={,\ }]{e1,e4}}
			\address[C]{Department of Statistics \& Actuarial Science, The University of Hong Kong, Hong Kong, China \printead[presep={,\ }]{e3}}
		\end{aug}
		
\begin{abstract}
	
	Small and medium-sized enterprises (SMEs) play a crucial role in driving economic growth. Monitoring their financial performance and discovering relevant covariates are essential for risk assessment, business planning, and policy formulation. This paper focuses on predicting profits for SMEs. Two major challenges are faced in this study: 1) SMEs data are stored across different institutions, and centralized analysis is restricted due to data security concerns; 2) data from various institutions contain different levels of missing values, resulting in a complex missingness issue. To tackle these issues, we introduce an innovative approach named \emph{Vertical Federated Expectation-Maximization} (VFEM), designed for federated learning under a missing data scenario. We embed a new EM algorithm into VFEM to address complex missing patterns when full dataset access is unfeasible. Furthermore, we establish the linear convergence rate for the VFEM and establish a statistical inference framework, enabling covariates to influence assessment and enhancing model interpretability. Extensive simulation studies are conducted to validate its finite sample performance. Finally, we thoroughly investigate a real-life profit prediction problem for SMEs using VFEM. Our findings demonstrate that VFEM provides a promising solution for addressing data isolation and missing values, ultimately improving the understanding of SMEs' financial performance.	
\end{abstract}
		
		
	\end{frontmatter}
	%%%%%%%%%%%%%%%%%%%%%%%%%%%%%%%%%%%%%%%%%%%%%%
	%% Please use \tableofcontents for articles %%
	%% with 50 pages and more                   %%
	%%%%%%%%%%%%%%%%%%%%%%%%%%%%%%%%%%%%%%%%%%%%%%
	%\tableofcontents
	
	\section{Introduction}
\label{sec:intro}

Small and medium-sized enterprises (SMEs) are the backbone of the world's economy \citep{acs1997small, hillary2017small}.
They play a pivotal role in facilitating innovation, entrepreneurial endeavors, and employment opportunities, particularly within the local labor market \citep{bagale2021small, garcia2023untangling}.
However, there is a high probability for SMEs to end up with failure.
%Profitablity is one of the preconditions for long-term firm survival and success.
Therefore, monitoring the profitability of SMEs has vital implications for risk assessment, business planning, and policy formulation \citep{shah2013factors, serrasqueiro2023smes}.
Good prediction of profitability enables SMEs to keep abreast of their operating situations and to take effective countermeasures.
It also facilitates investors to understand the firm value and to make rational investment decisions.
Moreover, analyzing the factors affecting profitability enables policymakers to improve supportive ecosystems and facilitate access to resources to foster the success of SMEs.

\begin{figure}[htb]
	\centering
	\includegraphics[width=0.8\linewidth]{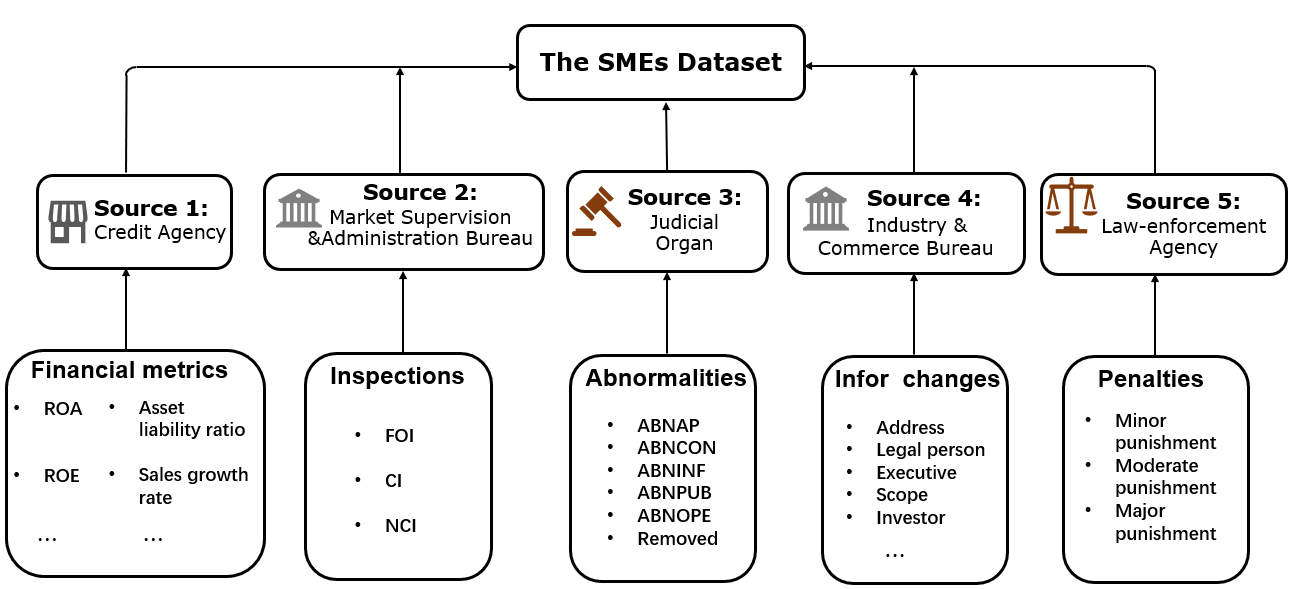}
	\caption{Five data sources containing information about SMEs.}
	\label{f3}
\end{figure}

To better predict the profitability of SMEs, it is necessary to utilize as much data as possible. Nevertheless, the relevant data for SMEs are often stored in different institutions; see Figure \ref{f3} for example. The \emph{Credit Agency} owns financial indicators of SMEs, and the \emph{Industry and Commerce Bureau} holds comprehensive commercial registration information. The \emph{Market Supervision and Administration Bureau} possesses detailed records regarding routine inspections and audits conducted on businesses. The \emph{Law-enforcement Agency} owns
administrative penalty records imposed on SMEs, while the \emph{Judicial Organ} provides litigation information. The profit prediction of SMEs can be greatly enhanced if data from these sources are well utilized.

Although there are multiple sources of data available, due to concerns over data privacy and security, data from different institutions cannot be shared freely. This poses a big challenge since it prevents us from entering the data together for comprehensive analysis. To address this issue, federated learning \citep{mcmahan2017communication} emerges as a promising solution. It is a machine learning approach that enables multiple data holders to collaborate in training models without the need for direct sharing of raw data. This privacy-preserving framework allows each data holder to keep its data local while contributing to the overall learning process. In general, federated learning can be broadly categorized into two types: horizontal and vertical. Horizontal federated learning (HFL) focuses on the case where multiple datasets share the same feature space but differ in the samples they contain. In contrast, vertical federated learning (VFL) handles datasets that have overlapping samples but distinct feature spaces. To handle the profit prediction task of SMEs using multiple sources as described in Figure \ref{f3}, the VFL framework is applicable, since different institutions store distinct features.

Due to its practical merits in facilitating data collaboration across multiple data sources, VFL has attracted increasing attention from both academia and industry. For instance, numerous Internet companies such as ByteDance, JD, and Tencent have embraced VFL to enhance their advertising efficiency \citep{cai2020bytedance, hy2021JD, ly2021tencent, wy2021Huawei}. Moreover, it shows promising applications in finance \citep{chen2021homomorphic} and healthcare \citep{cha2021implementing}. However, most traditional VFL approaches can only utilize fully observed samples. That is, the covariates for each sample are fully observed in all data sources. However, the SMEs dataset suffers from severe missing value issues. Just a small fraction of samples have all the covariates available in all institutions, while the remaining majority have only a subset of covariates observed in different institutions. This results in the missingness level of covariates in SMEs data varying from 0.91\% (for those in \emph{the Industry and Commerce Bureau}) to 93.28\% (for those in \emph{the Law-enforcement Agency}). This server missingness problem poses a big challenge to the profit prediction task.

In the past literature, there exists research focusing on predictive tasks involving missing values under the VFL setting, which can be roughly divided into two streams. The first stream of studies aims to fill in the missing values by imputation.
For example, \cite{du2024privacy} developed the $K$-nearest neighbors (KNN) imputation method which identifies $K$ most similar samples in each data set to impute the values of the missing data points. \cite{ren2024novel} designed an adaptive imputation module to refine imputation quality in an unsupervised manner, to address the issue of incomplete multi-view data across clients.
\cite{xin4775836tabular} proposed more sophisticated imputation methods based on conditional generative adversarial network (GAN).
The second stream of studies concentrates on studying representations for missing values. One specific situation is to predict pseudo-labels for unlabeled samples (regarded as missing values) to expand the training set. Some typical works include \cite{FENG2022118097, he2023hybrid} and \cite{li2023vfedssd}, which leveraged self-supervised learning to boost the representation learning capability of local models by exploiting unlabeled samples. \cite{FedCVT22} and \cite{YangYitao22} proposed semi-supervised learning approaches to augment labeled and fully observed samples.
\cite{zhang2022data, chen2023metadata} and \cite{xiao2024distributed} employed GAN to complete the representation of missing values by leveraging the features of limited aligned samples and abundant unaligned samples.

Previous works mainly borrow strength from deep learning models, which demonstrate superior prediction performance but fall short in examining the influence of covariates and lacking interpretability. However, in this work, predicting the profit of SMEs as well as identifying influential covariates are both of paramount importance. Based on these considerations, we adopt the linear regression model for help. Traditional approaches for handling missing values in linear regression models often apply the imputation methods, which aim to fill in the missing values by taking advantage of the available information. Classic imputation methods include single imputation, multiple imputation \citep{rubin2018multiple}, and multivariate imputation by chained equations (MICE) \citep{beesley2021stacked}; see \cite{little2019statistical} for more methods. Another commonly used method is the expectation-maximization (EM) algorithm \citep{DEMP1977, ding2016algorithm, balakrishnan2017statistical}. The EM algorithm iteratively imputes the missing values based on the observed data and then updates the regression parameters using the complete data. After convergence, the EM solution converges to the maximum likelihood estimator (MLE) under some mild conditions \citep{sundberg1974maximum, louis1982finding}. However, the traditional EM algorithm cannot be directly applied to the SMEs data, since certain quantities computed during the EM iterative process require access to the entire dataset, which is not feasible in the VFL setting.

Driven by the analysis of profitability for SMEs, we introduce an innovative approach named \emph{Vertical Federated Expectation-Maximization} (VFEM), which is designed for federated learning under a missing data scenario. VFEM enables local computation on data from each participating organization, requiring only the exchange of essential summary statistics. Moreover, a newly designed EM algorithm is embedded into VFEM to tackle intricate missing patterns when accessing the complete dataset is unattainable. Theoretically, we establish the linear convergence rate for the VFEM and establish a statistical inference framework, enabling covariates to influence assessment and enhancing model interpretability. To establish the inference framework, we propose a vertical sketching method for approximating the true asymptotic covariance matrix, avoiding the need to exchange original data. We conduct extensive simulation studies to demonstrate the finite sample performance of VFEM. Finally, we thoroughly investigate a real-life profit prediction problem for SMEs using VFEM. Compared with baseline methods, VFEM not only delivers superior prediction performance but also identifies key indicators that are relevant to the profitability of SMEs.

The rest of this work is organized as follows. In Section 2, we introduce the SMEs dataset in detail and conduct some preliminary analysis. We present the VFEM method in Section 3 and discuss its theoretical properties in Section 4. We apply VFEM to the SMEs dataset in Section 5. Concluding remarks and discussions are presented in Section 6. All technical proofs and simulation studies are present in the supplementary materials.

\section{The SMEs Data}

\subsection{Data Description}

In this work, we focus on the profit prediction of SMEs. To this end, we use five sources of information. They are, respectively, (1) financial information stored in the \emph{Credit Agency}, (2) the registration information stored in the \emph{Industry and Commerce Bureau}, (3) the administrative penalty information stored in the \emph{Law-enforcement Agency}, (4) the business anomaly information stored in the \emph{Judicial Organ}, and (5) the annual inspection information stored in the \emph{Market Supervision and Administration Bureau}. This leads to a total of 36 variables for 166,207 small and medium-sized enterprises engaged in various industries. Table \ref{table-d} summarizes the SMEs dataset, and the detailed description of variables can be found in Appendix D of the supplementary materials. The SMEs dataset has two notable characteristics.

\begin{table}[h]
	\centering
	\caption{Summaries of the SMEs dataset.}
	\footnotesize
	\tabcolsep=0.1cm
	\label{table-d}
	\begin{tabular}{@{}cccc@{}}
		\toprule
		Client index & Source of information & Number of variables & Missing proportion \\ \midrule
		1 & Credit Agency & 12 & 53.65\% \\
		2 & Market Supervision and Administration Bureau & 3 & 87.61\% \\
		3 & Judicial Organ & 6 & 93.05\% \\
		4 & Industry and Commerce Bureau & 9 & 0.91\% \\
		5 & Law-enforcement Agency & 5 & 93.28\% \\
		\bottomrule
	\end{tabular}
\end{table}

(1) Firstly, the dataset is compiled from five institutions. Owing to data security considerations, centralizing the data for comprehensive analysis is not feasible. To mitigate this challenge, the federated learning approach emerges as a viable solution, wherein each institution is regarded as an independent client. This approach ensures that data privacy is guaranteed while enabling collaborative analysis across multiple institutions.

(2) Secondly, the dataset is marked by the presence of missing values. Notably, these missing values are relatively comprehensive, signifying that if an enterprise lacks records within a particular institution, the entirety of records on that client will be absent. For example, different from large enterprises, many SMEs do not have their accounting systems, resulting in the financial information records in \emph{Credit Agency} being incomplete. In addition, those SMEs not incurring administrative penalties, do not have records in the \emph{Law-enforcement Agency}. Thus all covariates stored in \emph{Law-enforcement Agency} are not observed for those SMEs. Such missing pattern is also studied in \cite{wang2015merging}.
As illustrated in Table \ref{table-d}, the prevalence of missing values varies across different clients, with some exhibiting notably high rates. This complex missingness issue presents considerable challenges for subsequent data analysis since traditional missing data methods often focus on uniform missing rates.

\subsection{Preliminary Analysis}

To characterize the profitability and growth prospects of SMEs, we consider the \emph{net profit growth rate} (NPGR) as the response variable. It is a financial metric that quantifies the relative change in net profit between consecutive years about the net profit of the previous year. The other 35 variables are considered covariates, which characterize the operational performance, financial performance, credit records, legal litigations, and other relevant characteristics of these enterprises. During the data preprocessing stage, we carefully examined each continuous variable and found that some exhibited skewed distributions. To mitigate the influence of extreme values, we performed 1\% top and bottom truncation on the continuous covariates as long as the response variable. To explore the influences of covariates on NPGR, we adopt a linear regression model. However, as we discussed above, the analysis of the SMEs dataset faces two challenges. First, these covariates are stored in different institutions. Second, data in different institutions encounter different levels of missingness. As a result, only 96 samples are completely observed across all institutions in the SMEs dataset. To analyze the SMEs dataset under the two challenges, we explore three preliminary methods.

{\sc (1) The Single Method.} Considering the inability to share data across different institutions, our analysis solely relies on the dataset available in Client 1, where the response variable is stored. In light of the missing value issue, only the complete observed samples within Client 1 are utilized. We refer to this method as the Single method. The sample size in this method is 89,170.

{\sc (2) The CC Method.} We adopt the federated learning framework to leverage data from all institutions. However, in response to the missingness issue, we restrict our analysis to samples that are completely observed across all institutions. We refer to this method as the CC (complete case) method. The corresponding sample size is only 96.

{\sc (3) The Impute Method.} We also employ federated learning to analyze data from all institutions. However, to tackle the missingness issue, we initially impute the missing values using the mean of the observed ones and then build the linear regression model. We refer to this method as the Impute method. The sample size in this method is 166,207.

Table \ref{table-CI} presents the estimated coefficients and adjusted R squares of the three methods, from which we can draw the following conclusions. First, the Impute method has notably low R squares. Additionally, most coefficients in the Impute method exhibit different signs compared to other methods. This suggests that, due to the exceptionally high missing proportions in certain institutions, simply imputing the missing values performs poorly. Second, the Single method yields lower R squares than the CC method, implying the usefulness of information from other clients. Last, although the CC method yields considerable R squares, it fails to demonstrate statistical significance on the majority of covariates. For instance, the variables ASC (the amount of shareholder contribution), ROA, and Scale have been identified as significant to the profitability of SMEs in previous studies   \citep{valaskova2018financial, malakauskas2021financial}. However, under the CC method, these variables are not significant at the 5\% significance level. Based on the preliminary analysis we conclude that none of the aforementioned methods obtain desirable results. Hence it is necessary to develop a more effective federated learning approach that can better leverage missing values in the linear regression model.

\begin{table}[h]
	\centering
	\caption{The estimated coefficients and adjusted R squares of the Single, CC, and Impute methods on the SMEs dataset. The standard error is reported in parentheses, and * indicates the coefficient is significant under the 5\% significance level.}
	\scriptsize
	\renewcommand\arraystretch{1.2}
	\tabcolsep=0.1cm
	\label{table-CI}
	\begin{tabular}{@{}ccccc@{}}
		\toprule
		Client &	Variable & Single & CC & Impute \\ \midrule
		\multirow{12}*{\tabincell{c}{Credit Agency}}  &	EMP & 0.0019 (0.0008)* & 0.0681 (0.0327)*  &  -0.0339 (0.0018)*   \\
		&	ASC &	0.0018 (0.0006)* & 0.6529 (0.6573)	&  -0.5969 (0.0204)*  \\
		&	ROA &	-0.0260 (0.0054)* & 0.5705 (0.3273)	&  1.4147 (0.0126)*   \\
		&	ROE &	-0.0146 (0.0034)*   & 0.8366 (0.1932)*	&  -0.3825 (0.0083)*  \\
		&	OPM &	-0.0148 (0.0025)*  & -0.7585 (0.6532)  &  -0.4246 (0.0198)*  \\
		&	Scale &	-0.0569 (0.0022)* & 0.0741 (0.0755)	&  -0.0755 (0.0034)*   \\
		&	TBR   &	0.0035 (0.0008)* & -0.0136 (0.0401)  &  0.1222 (0.0018)*   \\
		&	ALR &	-0.2467 (0.0208)*  & -0.5904 (0.0997)* &  0.0429 (0.0047)*   \\
		&	ER  &	0.0434 (0.0069)*  & 0.0675 (0.0311)*	&  0.0083 (0.0014)*   \\
		&	SGR &	0.0562 (0.0006)*  & 0.0235 (0.0361)   &  0.0292 (0.0015)*   \\
		&	EGR &	0.0395 (0.0010)*  & 0.0043 (0.0621)	&  0.0235 (0.0022)*   \\
		&	AGR &	0.0051 (0.0011)*  & 0.1668 (0.0642)*	&  -0.0677 (0.0024)*  \\ \midrule
		\multirow{3}*{\tabincell{c}{Market Supervision and\\ Administration Bureau}}
		&	FOI &	-- & 1.0894 (0.4355)*	&  -0.1472 (0.0940)   \\
		&	CI  &	-- & -0.9192 (0.5190)   &   0.0374 (0.0977)   \\
		&	NCI &	-- & -0.9650 (0.5868)  &	0.6789 (0.1157)*  \\ \midrule
		\multirow{6}*{\tabincell{c}{Judicial Organ}} &	ABNAP  &	--  & -0.5648 (0.4349)   & -0.0465 (0.0553)\\
		&	ABNCON &	--  & -1.5547 (0.7141)* & 0.0515 (0.1154)  \\
		&	ABNINF &  --    & 1.1850 (0.6984)   & -2.6257 (0.1812)* \\
		&	ABNPUB & --     & -1.0497 (1.4638)  & 1.0483 (0.3778)*   \\
		&	ABNOPE & --     & -2.4197 (4.2349)  & -1.3478 (0.8776)  \\
		&	Removed & --    & -0.6018 (0.6071) & -0.5871 (0.1023)* \\ \midrule
		\multirow{9}*{\tabincell{c}{Industry and  \\Commerce Bureau}}
		&	ADDR & --  & -0.6046 (0.7064)    & 0.9804 (0.0323)*  \\
		&	LP   & --  & 0.4368 (0.6505)	   & 0.0323 (0.0354)   \\
		&	EXECU & --  & 1.1437 (0.7450)	   & -1.1334 (0.0401)* \\
		&	SCOPE & --  & -0.1881 (0.4515)   & -1.6775 (0.0239)* \\
		&	COMNAME  & -- & -0.1321 (1.2006) & 0.7315 (0.0509)* \\
		&	TERM  & --  & 1.4546 (0.6877)*	   & 1.6881 (0.0381)*  \\
		&	INVE  & --  & 2.8771 (0.5757)*	   & -0.7099 (0.0323)* \\
		&	CAPITAL & --  & -0.1763 (0.7884) & 0.0961 (0.0361)*  \\
		&	CHANGES & --  & -0.2035 (0.2008) & 0.6296 (0.0096)*  \\ \midrule
		\multirow{5}*{\tabincell{c}{Law-enforcement Agency}}
		&	FAP & -- & 0.0829 (0.0122)	   & -0.8168 (0.0020)*  \\
		&	FP1 & -- & -0.0253 (1.3570)     & -0.7080 (0.1859)*  \\
		&	FP2 & -- &  -0.8555 (0.7514)    & -0.5190 (0.1158)*  \\
		&	FP3 & -- &   0.1968 (0.8237)	   & 2.3565 (0.1498)*   \\
		&	Fines  & --& -0.0069 (0.5560)   & 0.0181 (0.0900)*  \\ \midrule
		Adjusted $R^2$ & -- & 0.2537 & 0.4348     & 0.0138    \\
		\bottomrule
	\end{tabular}
\end{table}

\section{The VFEM Methodology}\label{sec2}
\subsection{Model and Notations}

Assume we have $(\mathbf{x}_i,y_i)$ with $1\leq i \leq n$, where $y_i \in \mathbb{R}^1$ is a continuous response variable (e.g.,  NPGR in the SMEs data), and $\mathbf{x}_i \in \mathbb{R}^p$ is the centralized $p$-dimensional covariates. Assume the $p$-dimensional
covariates are stored in a total of $K$ clients (e.g., five institutions in the SMEs data). That is, $\mathbf{x}_i \triangleq (\mathbf{x}_i^{1\top}, \ldots, \mathbf{x}_i^{K\top})^{\top}$, where $\mathbf{x}_i^k \in \mathbb{R}^{p_k}$ is the covariate vector stored on the $k$-th client and $p_k$ is the associated dimension. Accordingly, we have $p = \sum_{k=1}^{K}p_k$. Without loss of generality, assume the response variable is stored in the first client (e.g., the \emph{Credit Agency}). Then in vertical federated learning, the first client is also referred to as the ``server''. The server plays a pivotal role in coordinating and managing the communication of information across different clients, as well as model aggregation among the participants. Let $\mathbf{Y} = (y_1, \ldots, y_n)^{\top}$ denote the response vector, and $\mathbf{X}_k = (\mathbf{x}^k_1, \ldots, \mathbf{x}^k_n)^{\top}$ the covariate matrix specific to the $k$-th client. The whole dataset $D$ can be represented as $D = D_1 \cup \ldots \cup D_K$, where $D_1 := (\mathbf{X}_1, \mathbf{Y})$ and $D_k := \mathbf{X}_k$ with $2 \leq k \leq K$.

\begin{figure}[htb]
	\centering
	\includegraphics[width=0.9\linewidth]{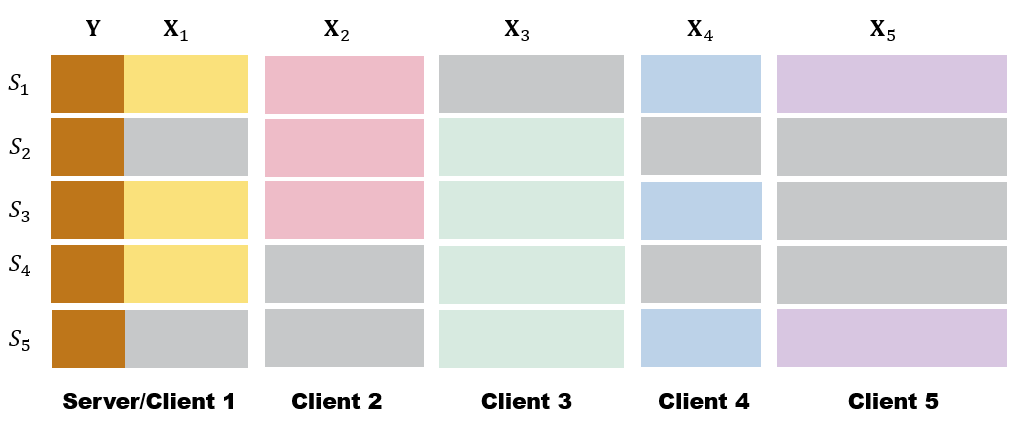}
	\caption{Illustration of vertically federated dataset with missing covariates. The dataset with $n$ samples and $p$ covariates is distributed among five clients. Here $S_1,\ldots, S_5$ are the index set of samples satisfying $S_1 \cup \ldots \cup S_5 = \{1,\ldots,n\}$ and $S_1 \cap \ldots \cap S_5 = \emptyset$. The colored and grey blocks indicate observed and missing values, respectively.}
	\label{f1}
\end{figure}

Suppose the response vector $\mathbf{Y}$ is fully observed by the server, but the $p$-dimensional covariate $\mathbf{x}_i$ for $1\leq i \leq n$ suffers from the missing problem. To align with the SMEs data, we consider a specific missing pattern in this work. That is, if the missingness happens for sample $i$ on client $k$, then the whole covariate vector $\mathbf{x}_i^k$ on client $k$ is missing. In other words, the client has no information about sample $i$ at all.
Specifically,
for each sample $i$, define the missing indicator vector as $M_i = (M_{i1}, \ldots, M_{iK})^{\top}$, where $M_{ik} = 1$ if the $p_k$-dimensional covariates on the $k$-th client are missing and $M_{ik} = 0$ otherwise. Thereby the matrix $M = (M_1, \ldots, M_n)^{\top}$ defines the missing data pattern. Define $\bm{\Delta}_{\text{obs}}$ as the index set of clients whose covariates are always observed, and $\bm{\Delta}_{\text{mis}}$ as the index set of clients encountering the missing issue. Further denote $\Delta_{i,\text{obs}}$ and $\Delta_{i, \text{mis}}$ to be the sets of clients where the corresponding covariates are observed or missing for sample $i$. That is, $\Delta_{i,\text{obs}} = \{1\leq k \leq K: M_{ik} = 0\}$ and $\Delta_{i,\text{mis}} = \{1\leq k \leq K: M_{ik} = 1\}$. Accordingly, we can rewrite
$\mathbf{x}_i = (\mathbf{x}_{i, \text{obs}}, \mathbf{x}_{i,\text{mis}})$, where $\mathbf{x}_{i,\text{obs}} = (\mathbf{x}_i^{k})$ with $k \in \Delta_{i,\text{obs}}$ and $\mathbf{x}_{i,\text{mis}} = (\mathbf{x}_i^{k})$ with $k \in \Delta_{i,\text{mis}}$. Note that the dimensions of $\mathbf{x}_{i,\text{obs}}$ and $\mathbf{x}_{i,\text{mis}}$ generally vary with sample $i$. For illustration purpose, Figure \ref{f1} presents a toy data example with five clients, where the colored and grey blocks indicate observed and missing values, respectively.

To model the relationship between $y_i$ and $\mathbf{x}_i$, assume the following regression model:
\begin{equation}\label{eq1}
	y_i = \sum_{k=1}^{K}\mathbf{x}_{i}^{k\top}\bm{\beta}_{k} + \epsilon_i,
\end{equation}
where $\epsilon_i$ is a random error independent of $\mathbf{x}_{i}^{k}$ for $1 \leq k \leq K$, and $\bm{\beta}_k \in \mathbb{R}^{p_k}$ is the regression coefficient vector on client $k$.
Let $\bm{\beta} = \left(\bm{\beta}_{1}^{\top}, \ldots, \bm{\beta}_{K}^{\top}\right)^{\top} \in \mathbb{R}^p$. Then $\bm{\beta}$ represents the total regression coefficients. Further assume $\epsilon_i \overset{\text { i.i.d. }}{\sim} \mathcal{N}(0, \sigma^2)$,
where $\mathcal{N}(\cdot)$ is the normal distribution.
Assume $\mathbf{x}_i^k \overset{\text { i.i.d. }}{\sim} \mathcal{N}_{p_k}(\bm{\mu}_k, \Sigma_k)$ with $1 \leq k \leq K$ and $1 \leq i \leq n$, where $\bm{\mu}_k$ and $\Sigma_k$ are mean vector and covariance matrix, respectively. Suppose the covariates across different clients are uncorrelated.
Denote $\bm{\theta} = \{\bm{\beta}, \bm{\mu}_1, \ldots, \bm{\mu}_K, \Sigma_1, \ldots, \Sigma_K, \sigma^2\}$ to be the collection of parameters to be estimated.
If the covariates could be fully observed, the complete-data log-likelihood is given by
\begin{equation}\label{eq2}
	\mathcal{L}(\bm{\theta}; \mathbf{X}, \mathbf{Y}) = \sum_{i=1}^{n} \mathcal{L}(\bm{\theta};\mathbf{x}_i, y_i) = \sum_{i=1}^{n}\left(\log (p(y_i|\mathbf{x}_i; \bm{\beta}, \sigma^2)) + \sum_{k=1}^{K}\log (p(\mathbf{x}_i^k; \bm{\mu}_k, \Sigma_k))\right),
\end{equation}
where $\mathbf{X} = (\mathbf{X}_1, \ldots, \mathbf{X}_K)$ is the complete covariate matrix. Then we can readily obtain the maximum likelihood estimator (MLE) through maximizing the complete-data log-likelihood \eqref{eq2}.
However, $\mathbf{X}$ has missing values. Then our objective is the observed-data log-likelihood \eqref{eq3}, which is obtained by marginalizing over the unobserved covariates.
\begin{equation}\label{eq3}
	\begin{split}
		\mathcal{L}(\bm{\theta};\mathbf{X}_{\text{obs}}, \mathbf{Y}) &= \sum_{i=1}^{n}\log \left(p(\mathbf{x}_{i,\text{obs}}, y_i; \bm{\theta})\right) \\
		&= \sum_{i=1}^{n}\log \left(\int p(\mathbf{x}_{i,\text{obs}}, y_i|\mathbf{x}_{i,\text{mis}}; \bm{\theta})p(\mathbf{x}_{i,\text{mis}};\bm{\theta})d \mathbf{x}_{i,\text{mis}}\right),
	\end{split}
\end{equation}
where $\mathbf{X}_{\text{obs}} = (\mathbf{x}_{1,\text{obs}}, \ldots, \mathbf{x}_{n,\text{obs}})^{\top}$ is the observed covariate matrix.
It is difficult to directly evaluate $\mathcal{L}(\bm{\theta};\mathbf{X}_{\text{obs}}, \mathbf{Y})$ due to the presence of missing covariates. Instead, we adopt the EM algorithm \citep{DEMP1977} to derive the MLE of \eqref{eq3}.
Below, we first discuss model estimation using EM by assuming all data are stored in one client in Section \ref{ssec22}, and then discuss the implementation of VFEM in Section \ref{ssec23}.

\subsection{Model Estimation}\label{ssec22}

Assume all data are stored in one client. We adopt the EM algorithm for model estimation, which is a popular technique to address the estimation problem with missing covariates.
The EM algorithm is composed of two steps. The first one is the E-step, where an auxiliary function $Q$ is computed. The second one is the M-step, which estimates model parameters by maximizing the $Q$-function.
Generally, the $Q$-function is defined as the expected complete log-likelihood with respect to the conditional distribution of the missing covariates, given the observed data and a current estimate of $\bm{\theta}$. Then the EM algorithm alternates between the E-step and M-step.

We first focus on the E-step. Recall that $\mathbf{x}_i$ is normally distributed. Then the conditional distribution of the missing covariates is
$\mathbf{x}_{i,\text{mis}}|\mathbf{x}_{i,\text{obs}},y_i \sim \mathcal{N}_{q_i}(\bm{\mu}_{i,\text{mis}|\text{obs}}, \Sigma_{i,\text{mis}|\text{obs}})$, where $q_i$ is the dimension of $\mathbf{x}_{i,\text{mis}}$. By simple calculations,
the conditional mean and conditional covariance of missing elements for sample $i$ can be derived as
\begin{equation}\label{neweq1}
	\begin{aligned}
		\bm{\mu}_{i,\text{mis}|\text{obs}} &=\bm{\mu}_{i,\text{mis}} + \left(\bm{\beta}_{i,\text{mis}}^{\top}\bm{\Sigma}_{i,\text{mis}}\bm{\beta}_{i,\text{mis}} + \sigma^2\right)^{-1} \bm{\Sigma}_{i,\text{mis}}\bm{\beta}_{i,\text{mis}}(y_i-\mu_{y_i}),\\
		\Sigma_{i,\text{mis}|\text{obs}} &= \Sigma_{i,\text{mis}} - \left(\bm{\beta}_{i,\text{mis}}^{\top}\Sigma_{i,\text{mis}}\bm{\beta}_{i,\text{mis}} + \sigma^2\right)^{-1} \Sigma_{i,\text{mis}}\bm{\beta}_{i,\text{mis}} \left(\Sigma_{i,\text{mis}}\bm{\beta}_{i,\text{mis}}\right)^{\top},
	\end{aligned}
\end{equation}
where $\mu_{y_i} = \mathbf{x}_{i,\text{obs}}^{\top}\bm{\beta}_{i,\text{obs}} + \bm{\mu}_{i,\text{mis}}^{\top}\bm{\beta}_{i,\text{mis}}$,
$\bm{\mu}_{i,\text{mis}} = (\bm{\mu}_k^{\top})^{\top}$, $\Sigma_{i, \text{mis}} = \operatorname{diag}(\Sigma_k)$, and $\bm{\beta}_{i,\text{mis}} = (\bm{\beta}_k^{\top})^{\top}$ with $k \in \Delta_{i,\text{mis}}$. We can define $\bm{\mu}_{i,\text{obs}}$, $\Sigma_{i, \text{obs}}$, and $\bm{\beta}_{i,\text{obs}}$ similarly.
Thus the $Q$-function is defined as the conditional expectation of the complete log-likelihood \eqref{eq2} given the current estimator $\widehat{\bm{\theta}}^{(t)}$, i.e.,
\begin{equation*}
	Q(\bm{\theta}|\widehat{\bm{\theta}}^{(t)}) = \frac{1}{n}\sum_{i=1}^{n}\int\left(\log (p(y_i|\mathbf{x}_i; \bm{\beta}, \sigma^2)) + \sum_{k=1}^{K}\log (p(\mathbf{x}_i^k; \bm{\mu}_k, \Sigma_k))\right)p(\mathbf{x}_{i,\text{mis}}|\mathbf{x}_{i,\text{obs}},y_i; \widehat{\bm{\theta}}^{(t)})d \mathbf{x}_{i,\text{mis}}.
\end{equation*}
Denote $\widetilde{\mathbf{x}}_i^k := \mathbf{x}^k_{i,\text{obs}}\mathbb{I}\left\{k \in \Delta_{i,\text{obs}}\right\}  + \bm{\mu}^k_{i,\text{mis}|\text{obs}}\mathbb{I}\left\{k \in \Delta_{i,\text{mis}}\right\}$, where $\mathbb{I}\{\cdot\}$ is the indicator function. Next, we define the pseudo-complete dataset $\widetilde{\mathbf{X}}_k := (\widetilde{\mathbf{x}}_1^k, \ldots, \widetilde{\mathbf{x}}_n^k)^{\top}$ for $k \in \bm{\Delta}_{\text{mis}}$. Specifically,
let $\widetilde{\mathbf{x}}_i = (\widetilde{\mathbf{x}}_i^{1\top}, \ldots, \widetilde{\mathbf{x}}_i^{K\top})^{\top}$, which represents imputing the missing entries of $\mathbf{x}_i$ with its corresponding conditional mean for sample $i$ ($1 \leq i \leq n$). Then with $\widetilde{\mathbf{x}}_i$, the explicit form of the $Q$-function is described as follows,
\begin{equation}\label{eq4}
	\begin{aligned}
		Q(\bm{\theta}|\widehat{\bm{\theta}}^{(t)}) &:= -\frac{1}{2}\log \sigma^{2} - \frac{1}{2}\sum_{k=1}^{K}\log|\Sigma_k| -\frac{1}{2n\sigma^{2}}\sum_{i=1}^{n}\left[y_i - \left(\beta_0 + \sum_{k=1}^{K} \widetilde{\mathbf{x}}_i^{k(t)\top} \bm{\beta}_{k}\right)\right]^2 \\
		& -\frac{1}{2n\sigma^{2}}\sum_{i=1}^{n}\bm{\beta}^{
			\top}_{i,\text{mis}}\Sigma^{(t)}_{i,\text{mis}|\text{obs}}\bm{\beta}_{i,\text{mis}}
		-\frac{1}{2n}\sum^K_{k=1}\sum_{i=1}^{n}\left(\widetilde{\mathbf{x}}_i^{k(t)} - \bm{\mu}_k\right)^{\top}\Sigma_{k}^{-1}\left(\widetilde{\mathbf{x}}_i^{k(t)} - \bm{\mu}_k\right)\\
		&-\frac{1}{2n}\sum^K_{k=1}\sum_{i=1}^{n}\operatorname{tr}\left(\Sigma_{k}^{-1} \Sigma^{k(t)}_{i,\text{mis}|\text{obs}}\right)\mathbb{I}\left\{k \in \Delta_{i,\text{mis}}\right\},
	\end{aligned}
\end{equation}
where $\Sigma^{k(t)}_{i,\text{mis}|\text{obs}} = \widehat{\Sigma}_k^{(t)} - \left(\widehat{\bm{\beta}}_{i,\text{mis}}^{(t)\top}\widehat{\Sigma}^{(t)}_{i,\text{mis}}\widehat{\bm{\beta}}^{(t)}_{i,\text{mis}} + \widehat{\sigma}^{2(t)}\right)^{-1} \widehat{\Sigma}^{(t)}_k\widehat{\bm{\beta}}^{(t)}_{k} \left(\widehat{\Sigma}^{(t)}_k\widehat{\bm{\beta}}^{(t)}_{k}\right)^{\top}$ with $k \in \Delta_{i,\text{mis}}$.

We then focus on the M-step, which updates $\widehat{\bm{\theta}}$ by maximizing \eqref{eq4}. Let $\widetilde{\mathbf{X}}^{(t)} = (\widetilde{\mathbf{x}}_1^{(t)},\ldots,\widetilde{\mathbf{x}}_n^{(t)})^{\top}$.
Then by maximizing \eqref{eq4}, we can easily obtain the closed form of parameters as follows,
\begin{equation}\label{eq5}
	\begin{aligned}
		& \widehat{\bm{\beta}}^{(t+1)} = \left(\widetilde{\mathbf{X}}^{(t)\top}\widetilde{\mathbf{X}}^{(t)} + \sum_{i=1}^{n}\widetilde{\Sigma}^{(t)}_{i,\text{mis}|\text{obs}}\right)^{-1}\widetilde{\mathbf{X}}^{(t)\top}\mathbf{Y},\\
		& \widehat{\bm{\mu}}_k^{(t+1)} = \frac{1}{n}\sum_{i=1}^{n}\widetilde{\mathbf{x}}_i^{k(t)}, \\
		& \widehat{\Sigma}_k^{(t+1)} = \frac{1}{n}\sum_{i=1}^{n}\left[(\widetilde{\mathbf{x}}_i^{k(t)} - \widehat{\bm{\mu}}_k^{(t)})^{\top}(\widetilde{\mathbf{x}}_i^{k(t)} - \widehat{\bm{\mu}}_k^{(t)}) + \Sigma^{k(t)}_{i,\text{mis}|\text{obs}}\mathbb{I}\{k \in \Delta_{i,\text{mis}}\} \right],\\
		& \widehat{\sigma}^{2(t+1)} = \frac{1}{n}\sum_{i=1}^{n}\left[\left(y_i - \sum_{k=1}^{K}\widetilde{\mathbf{x}}_i^{k(t)\top}\widehat{\bm{\beta}}^{(t)}_k\right)^2 + \widehat{\bm{\beta}}_{i,\text{mis}}^{(t)\top}\Sigma^{(t)}_{i,\text{mis}|\text{obs}}\widehat{\bm{\beta}}^{(t)}_{i,\text{mis}}\right],
	\end{aligned}
\end{equation}
where $\widetilde{\Sigma}^{(t)}_{i,\text{mis}|\text{obs}}$ is a $p \times p$ block matrix with its $(j, j^{\prime})$-th block being the corresponding block of $\Sigma^{(t)}_{i,\text{mis}|\text{obs}}$ if $j, j^{\prime} \in \Delta_{i,\text{mis}}$, otherwise being $\mathbf{0}_{p_j \times p_{j^{\prime}}}$.

\subsection{Federated Implementation}\label{ssec23}

To compute the explicit solutions in \eqref{eq5}, the population mean of covariates $\bm{\mu}_k$ on each client could be updated independently. However, the updates of other parameters rely on both the raw data (i.e., $\mathbf{x}_{i,\text{obs}}^k, y_i$) and local estimates (i.e., $\widehat{\Sigma}_k, \widehat{\bm{\beta}}_k, \widehat{\sigma^2}$) from different clients.
Unfortunately, the raw data are stored in different clients and prohibited to share due to privacy protection. This leads to the explicit solutions \eqref{eq5} practically infeasible.
To fix this problem,
we propose a vertical federated EM (VFEM) algorithm that can work well in a distributed manner without revealing the information of the raw data.

To adapt the classical EM algorithm to a federated learning setup,
we conduct an approximate maximization through a gradient ascent step in the M-step, which is also referred to as the first-order EM algorithm \citep{balakrishnan2017statistical}.
Based on some initialization values $\widehat{\bm{\theta}}^{(0)}$ and a learning rate $\eta \geq 0$, the updating formula in the M-step is given as:
\begin{equation*}
	\widehat{\bm{\theta}}^{(t+1)} = \widehat{\bm{\theta}}^{(t)} + \eta \left.\nabla Q\left(\bm{\theta} \mid \widehat{\bm{\theta}}^{(t)}\right)\right|_{\bm{\theta}=\widehat{\bm{\theta}}^{(t)}},
\end{equation*}
where $\nabla Q(\bm{\theta} \mid \widehat{\bm{\theta}}^{(t)})$ is the first-order derivate of the $Q$-function with respect to $\bm{\theta}$.

We first focus on the regression coefficients $\bm{\beta}$. Recall that $\bm{\beta}_k$ is the regression coefficient for covariates on client $k$. By easy calculations, we have
\begin{equation}\label{eq6}
	\left.\nabla_{\bm{\beta}_k} Q\left(\bm{\theta} \mid \widehat{\bm{\theta}}^{(t)}\right)\right|_{\bm{\theta}=\widehat{\bm{\theta}}^{(t)}} =
	\frac{1}{n}\sum_{i=1}^{n}\left[e_i^{(t)}\widetilde{\mathbf{x}}_i^{k(t)} - \bm{\alpha}^{(t)}_{i[k]}\mathbb{I}\left\{k \in \Delta_{i, \text{mis}}\right\}\right],
\end{equation}
where $e_i^{(t)} = y_i - \sum_{k=1}^{K} \widetilde{\mathbf{x}}_i^{k(t)\top} \widehat{\bm{\beta}}_{k}^{(t)}$ denotes the regression residual, $\bm{\alpha}^{(t)}_{i}=\Sigma^{(t)}_{i,\text{mis}|\text{obs}}\widehat{\bm{\beta}}^{(t)}_{i,\text{mis}}$ is a vector of dimension $q_i$, and $\bm{\alpha}^{(t)}_{i[k]}$ denotes the elements corresponding to client $k$ whose covariates are missing for individual $i$. Then the regression coefficient on client $k$ could be updated locally as
\begin{equation}\label{eq7}
	\widehat{\bm{\beta}}^{(t+1)}_k = \widehat{\bm{\beta}}^{(t)}_k + \eta \left.\nabla_{\bm{\beta}_k} Q\left(\bm{\theta} \mid \widehat{\bm{\theta}}^{(t)}\right)\right|_{\bm{\theta}=\widehat{\bm{\theta}}^{(t)}}.
	%	\hat{\beta}_0^{(t+1)} = \hat{\beta}_0^{(t)} + \eta \frac{1}{n}\sum_{i=1}^n e_i^{(t)}.
\end{equation}
Note that the gradients $\nabla_{\bm{\beta}_k} Q\left(\widehat{\bm{\theta}}^{(t)} \mid \widehat{\bm{\theta}}^{(t)}\right)$ consists of three parts: the regression residuals $e_i^{(t)}$, the local covariates themselves $\widetilde{\mathbf{x}}_i^{k(t)}$, and the parameter components $\bm{\alpha}_{i[k]}^{(t)}$ for $1 \leq i \leq n$.
In federated learning, although the covariates $\widetilde{\mathbf{X}}^{(t)}_k$ and response vector $\mathbf{Y}$ are not allowed to share, it is acceptable to share a summarized, noninvertible function of the raw data (e.g., $e_i^{(t)}, \bm{\alpha}^{(t)}_{i}, 1\leq i \leq n$).
Thus, to compute the local gradients \eqref{eq6}, we need to securely calculate $\widetilde{\mathbf{x}}_i^{k(t)}$,  $e_i^{(t)}$, and
$\bm{\alpha}_{i[k]}^{(t)}$ with $1 \leq i \leq n$. The computation details are given below.

\textbf{(1) Computation of $\widetilde{\mathbf{x}}_i^{k(t)}$}

For $\widetilde{\mathbf{x}}_i^{k(t)}$, it is acquired by imputing the missing value of $\mathbf{x}_i^k$ with its conditional mean $\bm{\mu}^{k(t)}_{i, \text{mis}|\text{obs}}$.
Based on \eqref{neweq1},
we have
$\bm{\mu}^{k(t)}_{i, \text{mis}|\text{obs}} = \widehat{\bm{\mu}}_{k}^{(t)} + (\sum_{k=1}^{K}\widehat{\bm{\beta}}^{(t)\top}_{k}\widehat{\Sigma}^{(t)}_k\widehat{\bm{\beta}}^{(t)}_{k}\mathbb{I}\{k \in \Delta_{i, \text{mis}}\} + \widehat{\sigma}^{2(t)})^{-1}\widehat{\Sigma}_k^{(t)}\widehat{\bm{\beta}}_k^{(t)}(y_i - \mu_{y_i}^{(t)})$,
wherein
$\mu_{y_i}^{(t)} = \sum_{k=1}^{K}\overline{\mathbf{x}}_i^{k(t)\top}\widehat{\bm{\beta}}_k^{(t)}$, while $\overline{\mathbf{x}}_i^{k(t)}$ is defined as $\overline{\mathbf{x}}_i^{k(t)} := \mathbf{x}^k_{i,\text{obs}}\mathbb{I}\{k \in \Delta_{i,\text{obs}}\} + \widehat{\bm{\mu}}_k^{(t)}\mathbb{I}\{k \in \Delta_{i,\text{mis}}\}$.
Accordingly, we conduct the following steps to obtain $\widetilde{\mathbf{x}}_i^{k(t)}$.
\begin{itemize}
	\item[1)] For $k \in \bm{\Delta}_{\text{mis}}$, the $k$th client imputes the missing values with their estimated population means $\widehat{\bm{\mu}}_k^{(t)}$ to compute $\overline{\mathbf{x}}_i^{k(t)}$;
	\item[2)] For $1\leq k \leq K$, the $k$th client computes its local ``fit'' $\overline{h}_i^{k(t)} = \overline{\mathbf{x}}_i^{k(t)\top}\widehat{\bm{\beta}}_k^{(t)}$;
	\item[3)] For $k \in \bm{\Delta}_{\text{mis}}$, the $k$th client computes $v_1^{k(t)} = \widehat{\bm{\beta}}^{(t)\top}_{k}\widehat{\Sigma}^{(t)}_k\widehat{\bm{\beta}}^{(t)}_{k}$ and transmits it to the server;
	\item[4)] The server computes the aggregated values $v_{2i}^{(t)} = \sum_{k=1}^{K}v_1^{k(t)}\mathbb{I}\{k \in \Delta_{i, \text{mis}}\} + \widehat{\sigma}^{2(t)}$ and $v_{3i}^{(t)} = y_i - \sum_{k=1}^{K}\overline{h}_i^{k(t)}$, which are then broadcasted to all clients;
	\item[5)] Next, the $k$th client with $k \in \bm{\Delta}_{\text{mis}}$ computes the conditional mean as $\bm{\mu}^{k(t)}_{i,\text{mis}|\text{obs}} = \widehat{\bm{\mu}}^{(t)}_k + \widehat{\Sigma}^{(t)}_k\widehat{\bm{\beta}}^{(t)}_k v_{3i}^{(t)}/v_{2i}^{(t)}$.
\end{itemize}

By this way, $\widetilde{\mathbf{x}}^{k(t)}_i$ is obtained as $\widetilde{\mathbf{x}}_i^{k(t)} = \mathbf{x}_{i,\text{obs}}^k \mathbb{I}\{k \in \Delta_{i,\text{obs}}\} + \bm{\mu}^{k(t)}_{i, \text{mis}|\text{obs}}\mathbb{I}\{k \in \Delta_{i, \text{mis}}\}$, thus finishing the E-step.
Notably, in the E-step, clients should impute their missing values twice. The first one is to compute the conditional mean $\bm{\mu}^{k(t)}_{i,\text{mis}|\text{obs}}$, where the current estimate of the population mean $\widehat{\bm{\mu}}_k^{(t)}$ is served as the imputation. The second one is to impute each missing value by its conditional mean $\bm{\mu}^{k(t)}_{i,\text{mis}|\text{obs}}$, which constructs the pseudo-complete dataset $\widetilde{\mathbf{X}}_k^{(t)}$, serving as a fundamental component for the subsequent M-step.

\textbf{(2) Computation of $e_i^{(t)}$}

It is straightforward to compute the regression residuals $e_i^{(t)}$ collaboratively. For this purpose,
all clients just need to calculate an inner-product between the raw data and parameters, i.e., $\widetilde{h}_i^{k(t)} = \widetilde{\mathbf{x}}_i^{k(t)\top}\widehat{\bm{\beta}}_k^{(t)}, 1\leq i \leq n$, and then transmit them to the server. Next, the residuals are computed as $e_i^{(t)} = y_i - \sum_{k=1}^{K}\widetilde{h}_i^{k(t)}, 1\leq i \leq n$. Then the server sends residuals back to each client to update the local parameters.

\textbf{(3) Computation of $\bm{\alpha}_i^{(t)}$}

Recall $\bm{\alpha}_i^{(t)} = \Sigma^{(t)}_{i, \text{mis}|\text{obs}}\widehat{\bm{\beta}}^{(t)}_{i,\text{mis}}$, where $\Sigma^{(t)}_{i, \text{mis}|\text{obs}} = \widehat{\Sigma}^{(t)}_{i,\text{mis}} - (\sum_{k=1}^{K}\widehat{\bm{\beta}}^{(t)\top}_{k}\widehat{\Sigma}^{(t)}_k\widehat{\bm{\beta}}^{(t)}_{k}\mathbb{I}\{k \in \Delta_{i, \text{mis}}\} + \widehat{\sigma}^{2(t)})^{-1} \widehat{\Sigma}^{(t)}_{i,\text{mis}}\widehat{\bm{\beta}}^{(t)}_{i,\text{mis}}(\widehat{\Sigma}^{(t)}_{i,\text{mis}}\widehat{\bm{\beta}}^{(t)}_{i,\text{mis}})^{\top}$. The conditional covariance of missing entries $\Sigma^{(t)}_{i, \text{mis}|\text{obs}}$ requires collaborative computation. Thus the computation of $\bm{\alpha}_i^{(t)}$ is conducted through the following steps.
\begin{itemize}
	\item[1)] For $k \in \bm{\Delta}_{\text{mis}}$, the $k$th client should send to the server the intermediate vectors $\mathbf{v}_{1}^{k(t)} = \widehat{\Sigma}^{(t)}_k\widehat{\bm{\beta}}^{(t)}_k$;
	\item[2)] The server forms an intermediate vector $\mathbf{v}_{2i}^{(t)} = (\mathbf{v}_1^{k(t)\top})^{\top}$ with $k \in \Delta_{i, \text{mis}}$, and computes an intermediate matrix $\mathbf{V}_i^{(t)} = \mathbf{v}_{2i}^{(t)}\mathbf{v}_{2i}^{(t)\top}$ for sample $i$ with missing values;
	\item[3)] The intermediate matrix $\mathbf{V}_i^{(t)}$ is transferred to each client. Then each client computes $\mathbf{v}_{3i}^{k(t)} = \mathbf{V}_{i[\cdot k]}^{(t)}\widehat{\bm{\beta}}_k^{(t)}/v_{2i}^{(t)}$, where $\mathbf{V}_{i[\cdot k]}^{(t)}$ denotes the $p_k$ columns of the $(q_i \times q_i)$-dimentional matirx $\mathbf{V}_{i}^{(t)}$ with $k \in \Delta_{i, \text{mis}}$ for sample $i$;
	\item[4)] For $k \in \bm{\Delta}_{\text{mis}}$, the $k$th client reports $\mathbf{v}_{3i}^{k(t)}$ to the server. The server then calculates the final aggregation  as $\mathbf{v}_{4i}^{k(t)} = \sum_{k \in \Delta_{i, \text{mis}}}\mathbf{v}_{3i}^{k(t)}$, which is then broadcast back to the clients;
	\item[5)] For $k \in \bm{\Delta}_{\text{mis}}$, the $k$th client computes $\bm{\alpha}^{(t)}_{i[k]} = \mathbf{v}^{k(t)}_{1} - \mathbf{v}_{4i[k]}^{k(t)}$.
\end{itemize}

Based on $\widetilde{\mathbf{x}}^{k(t)}_i$, $e_i^{(t)}$, and $\bm{\alpha}^{(t)}_{i[k]}$ for $1\leq i \leq n$, local gradients \eqref{eq6} can be computed on each client. Then the coefficient $\bm{\beta}_k$ could be updated locally according to \eqref{eq7}. After convergence (e.g., using $T$ updating steps), we obtain $\widehat{\bm{\beta}}^{(T)}_k$, which is the VFEM estimator for $\bm{\beta}_k$.

The other parameters $\bm{\mu}_k$, $\Sigma_k$, and $\sigma^2$ are still updated as \eqref{eq5}.
Note that $\widehat{\Sigma}_k^{(t+1)}$ contains the statistic $\Sigma^{k(t)}_{i, \text{mis}|\text{obs}} = \widehat{\Sigma}^{(t)}_k - \mathbf{v}_1^{k(t)}\mathbf{v}_1^{k(t)\top}/v_{2i}^{(t)}$, where $v_{2i}^{(t)}$ and $\mathbf{v}_1^{k(t)}$
are computed when updating $\widehat{\bm{\beta}}_k$. Then each client is able to update the estimates of local population covariance $\Sigma_k$ independently. As for $\widehat{\sigma}^{2(t+1)}$, it requires compuation of the statistic
$v_{4i}^{(t)} = \widehat{\bm{\beta}}^{(t)\top}_{i,\text{mis}}\Sigma^{(t)}_{i, \text{mis}|\text{obs}}\widehat{\bm{\beta}}^{(t)}_{i,\text{mis}}$. To this end, each client needs to compute $v_{5i}^{k(t)} = \widehat{\bm{\beta}}_k^{(t)\top}\bm{\alpha}_{i[k]}^{(t)}$. Then the server collects $v_{5i}^{k(t)}$ and computes $v_{4i}^{(t)} = \sum_{k \in \Delta_{i, \text{mis}}}v_{5i}^{k(t)}$. Based on $v_{4i}^{(t)}$, the server could obtain the estimate of $\sigma^2$ as $\widehat{\sigma}^{2(t+1)} = \sum_{i=1}^{n}((e_i^{(t)})^2 + v_{4i}^{(t)})/n$.
The details of VFEM are summarized in Algorithm \ref{alg1}.

\begin{algorithm}[!h]
	\caption{Implementation of the VFEM Method}
	\label{alg1}
	\footnotesize
	\KwIn{Maximum iterative number $T$, convergence value $\epsilon$, learning rate $\eta$\, initial coefficient value $\widehat{\bm{\beta}}^{(0)}$, and
		initial distributional parameters $\widehat{\sigma}^{2(0)} = \frac{1}{n}\sum_{i=1}^{n}(y_i-\bar{Y})^2$ with $\bar{Y}=\frac{1}{n}\sum_{i=1}^{n}y_i$,
		$\widehat{\bm{\mu}}_k^{(0)}=\frac{1}{m_k}\sum_{i=1}^{n}\mathbf{x}^k_{i,\text{obs}}\mathbb{I}\{k \in \Delta_{i,\text{obs}}\}$,
		$\widehat{\Sigma}_k^{(0)} = \frac{1}{m_k}\sum_{i=1}^{n}(\mathbf{x}^k_{i,\text{obs}} - \widehat{\bm{\mu}}_k^{(0)})(\mathbf{x}^k_{i,\text{obs}} - \widehat{\bm{\mu}}_k^{(0)})^{\top}\mathbb{I}\{k \in \Delta_{i,\text{obs}}\}$, where $m_k$ is the number of completely observed cases on client $k$ for $1 \leq k \leq K$.}
	\KwOut{Estimated coefficient vector $\widehat{\bm{\beta}}$.}
	Let $t=0$\;
	\While{$|\ell^{(t+1)}-\ell^{(t)}|\geq \epsilon$ and $t \leq T$}{
		\textbf{E-step: estimate the missing covariates}\\
		\For{client $k =1,\cdots,K$}{
			computes $\overline{h}_i^{k(t)}, 1\leq i \leq n$ and $v_1^{k(t)}$, then uploads them to the server\;
		}
		The server computes $v_{2i}^{(t)}$ and $v_{3i}^{(t)}$ and broadcasts them to all clients\;
		\For{client $k =1,\cdots,K$}{
			computes the local conditional mean $\bm{\mu}^{k(t)}_{i,\text{mis}|\text{obs}}$ and obtain the pseudo-complete covariates $\widetilde{\mathbf{x}}_i^{k(t)}$\;
		}
		\textbf{M-step: update the model parameters}\\
		\textbf{1. For the coefficients}\\
		\For{client $k =1,\cdots,K$}{
			computes $\widetilde{h}_i^{k(t)}, 1\leq i \leq n$ and $\mathbf{v}_{1}^{k(t)}$, reports them to the server\;
		}
		The server computes residuals $e_i^{(t)}$ and $\mathbf{V}_i^{(t)}$ for $1\leq i \leq n$, broadcasting them to clients\;
		\For{client $k =1,\cdots,K$}{
			computes $\mathbf{v}_{3i}^{k(t)}, 1\leq i \leq n$, reports them to the server\;
		}
		The server calculates $\mathbf{v}_{4i}^{k(t)}, 1\leq i \leq n$, broadcasts them to clients\;
		\For{client $k =1,\cdots,K$}{
			computes $\bm{\alpha}_{i[k]}^{(t)}$ and calculates local gradients as \eqref{eq6}\;
			updates local coefficients as \eqref{eq7}\;
		}
		\textbf{2. For the distributional parameters}\\
		\For{client $k =1,\cdots,K$}{
			computes $\Sigma_{i,\text{mis}|\text{obs}}^{k(t)}$ and $v^{k(t)}_{5i}$ for $1 \leq i \leq n$\;
			updates $\widehat{\bm{\mu}}_k$ and $\widehat{\Sigma}_k$ as \eqref{eq5}\;
			transfers $v^{k(t)}_{5i}, 1 \leq i \leq n$ to the server\;
		}
		The server calculates $v_{4i}^{(t)},1\leq i\leq n$, utilizing them to update $\widehat{\sigma}^2$\;
		and computes the loss $\ell^{(t+1)} = \frac{1}{n}\sum_{i=1}^{n}\left((e_i^{(t)})^2 + v_{4i}^{(t)}\right)$\;
		$t = t+1$.
	}
\end{algorithm}

\section{Theoretical Analysis}

\subsection{The Convergence Property}

We first aim to study the convergence property of the VFEM estimator.
Denote the true regression coefficient as $\bm{\beta}^*$. For simplicity, we assume $\bm{\mu}_k = \mathbf{0}_{p_k}$ and $\operatorname{diag}(\Sigma_k) = \mathbf{1}_{p_k}$. Suppose the generative model \eqref{eq1} is correctly specified.
We further assume that the distributional parameters (i.e., $\bm{\mu}_k, \Sigma_k, \sigma^2$) are already known. Thus we only focus on the estimation of $\bm{\beta}^*$. Denote $\widehat{\bm{\beta}}^{(T)}$ to be the VFEM estimator, where $T$ is the total number of iterations of Algorithm \ref{alg1}. Before deriving the convergence property of $\widehat{\bm{\beta}}^{(T)}$, we first give some necessary assumptions.

\begin{assumption}\label{assum1}
	Assume the missing mechanism is missing completely at random (MCAR).
\end{assumption}	

\begin{assumption}\label{assum2}
	Let $\lambda_{min}/\lambda_{max}$ represent the minimum/maximum of the smallest/largest eigenvalues of the local covariance matrices $\Sigma_k$ with $1\leq k \leq K$. Define $\xi_1 := \|\bm{\beta}^*\|_2/\sigma$, $\xi_2 := r/\sigma$, where $r$ is the radius of the Euclidean ball $\mathbb{B}_2(r;\bm{\beta}^*)$. Define the probability of missingness on client $k$ as $\rho_k := \mathbb{P}(M_{ik}=1)$ for $1 \leq i \leq n, 1 \leq k \leq K$. Denote $\rho_{\max} := \max \{\rho_1,\ldots, \rho_K\}$, $\rho_{\min} := \min \{\rho_1,\ldots, \rho_K\}$,
	$\xi = \Big[\big(\lambda_{\max}(2\xi_1+\xi_2) + \lambda_{\min}\xi_2\big)\delta + \\ \lambda_{\max}\xi_2\Big] (\xi_1+\xi_2)$, and $\delta = \left(1+\left[\lambda_{\max}(\xi_1+\xi_2)^2\right]^{-1}\right)^{-1}$. Under the condition $\frac{\lambda_{\min}}{C\lambda_{\max}} > \delta$,
	$\rho_{\max}$ is assumed to satisfy the bound $0 < \rho_{\max} < \left(\frac{\lambda_{\min}}{C\lambda_{\max}}-\delta\right)/\xi$,
	where $C$ is a positive constant.
\end{assumption}

\begin{assumption}\label{assum3}
	Assume the sample size is lower bounded as $n \geq c_1p\log(1/\delta_1)$, where $c_1>0$ and $\delta_1>0$.	
\end{assumption}

For simplicity, we focus on the case where the missing occurs completely at random. That is, the missingness of the data is purely a result of chance or random fluctuations and is not influenced by any underlying factors or variables. This assumption is practically reasonable for missing data in most VFL applications. Similar assumptions are also
widely employed in the existing literature \citep{balakrishnan2017statistical, jin2021factor}. Although the MCAR assumption is necessarily needed in theoretical analysis, we investigate simulation studies beyond MCAR and find the estimation results are robust to the missing mechanism; see Appendix E in the supplementary materials for details. Assumption \ref{assum2} and Assumption \ref{assum3} provide essential requirements for the missingness probability $\rho_k$s and sample size $n$. Based on these assumptions, the convergence property of the VFEM estimator is summarized below.

\setcounter{theorem}{0}
\renewcommand\thetheorem{\arabic{theorem}}

\begin{theorem}\label{thm1} Assume Assumptions \ref{assum1}-\ref{assum3} hold. Denote $\gamma = C\lambda_{\max} \left[\rho_{\max} \xi+ (1-\rho_{\min})\delta\right]$, where $\xi$ and $\delta$ are defined in Assumption \ref{assum2}.
	Then for any initialization $\widehat{\bm{\beta}}^{(0)} \in \mathbb{B}_2\left(r;\bm{\beta}^*\right)$ and the stepsize $\eta = 2/(\lambda_{\max}+\lambda_{\min})$,  the VFEM estimator satisfies the bound
	\begin{equation}\label{eq14}
		\left\|\widehat{\bm{\beta}}^{(T)} - \bm{\beta}^*\right\|_2 \leq \underbrace{\left(1-\frac{2\left(\lambda_{\min}-\gamma\right)}{\lambda_{\max} + \lambda_{\min}}\right)^T \left\|\widehat{\bm{\beta}}^{(0)} - \bm{\beta}^*\right\|_2}_{\text{Optimization Error}}  + \underbrace{\frac{c_2 \lambda_{\max}^{3}\sigma^2}{\lambda_{\min}(\lambda_{\min}-\gamma)}\sqrt{\frac{p}{n}\log\left(\frac{1}{\delta_1}\right)}}_{\text{Statistical Error}} ,
	\end{equation}
	with probability at least $1-\delta_1$, where $\delta_1>0$ is a constant.
\end{theorem}

The proof of Theorem \ref{thm1} is given in Appendix A.3 in the supplementary materials.
Based on Theorem \ref{thm1}, we find that the upper bound of the overall estimation error \eqref{eq14} involves two terms. The first one is the upper bound of the optimization error, which decreases geometrically with the increase of the iteration number $T$ since we have $0<2(\lambda_{\min}-\gamma)/(\lambda_{\max}+\lambda_{\min}) < 1$. The second one is the upper bound of the statistical error, which is roughly of the order $\sqrt{p/n}$.
Therefore, Theorem \ref{thm1} guarantees that the VFEM estimator converges to a small neighborhood
of a given global optimum $\bm{\beta}^*$, given that the initialization falls into the region of attraction around this
fixed point. It is notable that, the size of the region of attraction is determined by the missing probabilities $\rho_{\min}/\rho_{\max}$ and the signal-to-noise ratio $\xi_1$. In general, increasing $\rho$ will cause the expanded size of the region. This makes $\widehat{\bm{\beta}}^{(T)}$ converge to a point that is beyond the desired distance to $\bm{\beta}^*$. However, $\widehat{\bm{\beta}}^{(0)} \in \mathbb{B}_2(r;\bm{\beta}^*)$ is only a technical assumption here. Our simulation studies in supplementary materials reveal that even a relatively poor initialization can make the VFEM estimator
converge to a near-globally optimal solution.

\subsection{The Asymptotic Property}

We study the asymptotic property of the VFEM estimator in this section. In general, the VFEM estimator should share the same asymptotic distribution as the classic EM estimator in the non-VFL setting. Here we only focus on the interested coefficient $\bm{\beta}$ and assume the nuisance parameters to be true. We first impose standard regularity assumptions which are commonly used in previous literature, and then present the asymptotic property of the VFEM estimator in Theorem \ref{thm4}.
\begin{assumption}\label{con1}
	The true coefficient $\bm{\beta}^*$ lies in the interior of a known compact set $\Omega_{\bm{\beta}}$.
\end{assumption}
\begin{assumption}\label{con2}
	The true $\bm{\beta}^*$ is the unique solution that maximizes $\mathbb{E}[\mathcal{L}(\bm{\beta}; \mathbf{X}_{\text{obs}}, \mathbf{Y})]$.
\end{assumption}
\begin{assumption}\label{con3}
	$\mathbb{E}_{\bm{\beta}^*}[-\partial^2 \mathcal{L}(\bm{\beta}; \mathbf{X}_{\text{obs}}, \mathbf{Y})/\partial \bm{\beta}]$ is strictly positive definite.
\end{assumption}
\begin{assumption}\label{con4}
	Uniformly for $\bm{\beta}$ in some neighborhood of $\bm{\beta}^*$, $\partial^3 \mathcal{L}(\bm{\beta}; \mathbf{X}_{\text{obs}}, \mathbf{Y})/\partial \bm{\beta}$ is dominated by some function which is integrable with respect to the density of the observed data.
\end{assumption}

\begin{theorem}\label{thm4}
	Assume Assumptions \ref{assum1}-\ref{con4} hold. Then as the sample size $n \rightarrow \infty$ and the iteration number $T \rightarrow \infty$, we have
	\begin{equation*}
		\sqrt{n}(\widehat{\bm{\beta}}^{(T)} - \bm{\beta}^*) \stackrel{d}{\longrightarrow} \mathcal{N}(\mathbf{0}_p, I(\bm{\beta}^*)),
	\end{equation*}
	where $I(\bm{\beta}^*) = - \mathbb{E}_{\bm{\beta}^*}\left[\nabla^2_{\bm{\beta},\bm{\beta}} \left.Q(\bm{\beta}|\widehat{\bm{\beta}}^{(t)})\right|_{\bm{\beta} = \bm{\beta}^*, \widehat{\bm{\beta}}^{(t)} = \bm{\beta}^*} + \nabla^2_{\bm{\beta}, \widehat{\bm{\beta}}^{(t)}} \left.Q(\bm{\beta}|\widehat{\bm{\beta}}^{(t)})\right|_{\bm{\beta} = \bm{\beta}^*, \widehat{\bm{\beta}}^{(t)} = \bm{\beta}^*}\right]$, and the expectation is taken with regard to both the distribution of the data and the missing mechanism. The explicit form of $Q(\bm{\beta}|\widehat{\bm{\beta}}^{(t)})$ is given in (A.1) in Appendix A in the supplementary materials.
\end{theorem}

The proof of Theorem \ref{thm4} is given in Appendix B in the supplementary materials. The intuition behind Theorem \ref{thm4} is straightforward. Given Assumptions \ref{assum1}-\ref{con4}, the VFEM estimator can numerically converge to the MLE of \eqref{eq3} in the non-VFL setting \citep{sundberg1974maximum, louis1982finding}, since the VFEM algorithm has nearly no loss of information compared to the classic EM algorithm. Therefore, similar to MLE, the VFEM estimator should also be guaranteed with asymptotic normality \citep{nielsen2000stochastic, wang2015high}. In Theorem \ref{thm4}, the information matrix $I(\bm{\beta}^*)$ could be obtained by taking the expectation of the second derivative of $Q(\bm{\beta}|\widehat{\bm{\beta}}^{(t)})$. However, $I(\bm{\beta}^*)$ is computationally intractable because it has no explicit formulation. Therefore, it is necessary to seek an estimate of the asymptotic covariance $I(\bm{\beta}^*)$ to facilitate statistical inference in practice.

\subsection{Computation of the Asymptotic Covariance}\label{sec33}

To start with, reformulate $\bm{\theta}$ as a $d$-dimensional vector $\bm{\theta} = (\bm{\beta}^{\top}, \bm{\mu}^{\top}, (\operatorname{vec}(\Sigma))^{\top}, \sigma^2)^{\top}$, where $\bm{\mu} = (\bm{\mu}_1^{\top}, \ldots, \bm{\mu}_K^{\top})^{\top}$, $\Sigma = \operatorname{diag}(\Sigma_1,\ldots,\Sigma_K)$, and $\operatorname{vec}(\cdot)$ means vectorization. Define $\widehat{\bm{\theta}}^{(T)}$ to be the VFEM estimator, and $\mathcal{\bm{V}}$ to be the
asymptotic covariance matrix of $\widehat{\bm{\theta}}^{(T)}$. According to \cite{meng1991using}, the estimated asymptotic covariance is
\begin{equation}\label{CI1}
	\mathcal{\bm{V}} = I_{oc}(\widehat{\bm{\theta}}^{(T)})^{-1}\left(\mathbf{I}_d - \Gamma \right)^{-1}.
\end{equation}
Here $I_{oc}(\widehat{\bm{\theta}}^{(T)}) = \left.\mathbb{E}\left(\left.- \nabla^2_{\bm{\theta}, \bm{\theta}}\mathcal{L}(\bm{\theta};\mathbf{X},\mathbf{Y}) \right|\mathbf{X}_{\text{obs}},\mathbf{Y},\bm{\theta}\right)\right|_{\bm{\theta} = \widehat{\bm{\theta}}^{(T)}}$ is the complete-data observed information matrix, whose explicit form is given in Appendix C.1 in the supplementary materials. $\mathbf{I}_d$ is the $d\times d$-dimensional identity matrix. Let $F(\bm{\theta}) = (F_1(\bm{\theta}),\ldots, F_d(\bm{\theta}))$ denote the mapping functions used in \eqref{eq5}. Then $\Gamma = \left(\partial F_j(\bm{\theta})/\partial \bm{\theta}_i)\right|_{\bm{\theta}=\widehat{\bm{\theta}}^{(T)}}$ represents the $d \times d$ Jacobian matrix for $F(\bm{\theta})$ evaluated at $\bm{\theta} = \widehat{\bm{\theta}}^{(T)}$. $\Gamma$ measures the component-wise rate of convergence of the VFEM algorithm, and is governed by the fraction of information loss due to incomplete observation \citep{DEMP1977, meng1991using, mclachlan2007algorithm}.

The computation of $\mathcal{\bm{V}}$ involves calculating $I_{oc}(\widehat{\bm{\theta}}^{(T)})$ and $\Gamma$. For $\Gamma$, we compute it using the numerical differentiation method \citep{meng1991using} in a federated manner. The detailed computational procedure is presented in Appendix C.2 in the supplementary materials. In terms of $I_{oc}(\widehat{\bm{\theta}}^{(T)})$, its computation can be easily conducted in the non-VFL setting. However, a significant challenge arises if the data are stored separately in different clients. This is because the computation of $I_{oc}(\widehat{\bm{\theta}}^{(T)})$ involves the term $\widetilde{\mathbf{X}}^{(T)}$, which is the pseudo-complete covariate matrix with the VFEM estimator $\widehat{\bm{\theta}}^{(T)}$ plugin. For example, by simple calculations, we have
$$I_{oc}(\widehat{\bm{\theta}}^{(T)})_{\bm{\beta}, \bm{\beta}} = -(\widetilde{\mathbf{X}}^{(T)\top}\widetilde{\mathbf{X}}^{(T)} + \sum_{i=1}^{n}\widetilde{\Sigma}^{(T)}_{i,\text{mis}|\text{obs}})/\widehat{\sigma}^{2(T)}.$$

To address this issue, we employ the sketching method \citep{woodruff2014sketching}. The basic idea of the sketching method is to conduct random projection. By applying random projection on the original data, it is feasible to acquire a condensed representation that retains the essential characteristics of the data. This reduced representation facilitates the generation of approximate solutions or estimates with an acceptable level of accuracy \citep{achlioptas2003database, halko2011finding, liu2023new}. Specifically, for each client $k$, we apply a random linear transformation to the original $(n\times p_k)$-dimensional covariate matrix $\widetilde{\mathbf{X}}^{(T)}_k$, which is formulated as $\bm{\mathcal{X}}_k = \mathbf{S}_k\widetilde{\mathbf{X}}^{(T)}_k$. Here
$\mathbf{S}_k = \mathbf{G}_k/\sqrt{m}$, and $\mathbf{G}_k \in \mathbb{R}^{m\times n}$ is the Gaussian sketching matrix with i.i.d. $\mathcal{N}(0,1)$ entries ($m \ll n$). Then the resulting $(m \times p_k)$-dimensional random representations $\bm{\mathcal{X}}_k$ are reported to the server. Denote the aggregated representations to be $\bm{\mathcal{X}} = (\bm{\mathcal{X}}_1,\ldots, \bm{\mathcal{X}}_K)$. Then the server can compute an approximate version of the original statistics. For example, let $\widehat{\mathbf{M}}_1 = \bm{\mathcal{X}}^{\top}\bm{\mathcal{X}}$. Then $\widehat{\mathbf{M}}_1$ is an estimate of $\mathbf{M}_1 =\widetilde{\mathbf{X}}^{(T)\top}\widetilde{\mathbf{X}}^{(T)}$. The other statistics can be calculated similarly.

To adjust for the additional variability induced by random approximation, we can repeat the sketching method for $L$
times in parallel and aggregate these results to restore statistical accuracy \citep{chen2016integrating, shen2023fadi}. Then the final estimate to $\mathbf{M}_1$ is $\widehat{\mathbf{M}}_1 = L^{-1}\sum_{l=1}^{L}\widehat{\mathbf{M}}_1^{(l)}$, where $\widehat{\mathbf{M}}_1^{(l)}$ is the corresponding result of the $l$-th random approximation. The detailed procedure is shown in Algorithm \ref{alg2}. Furthermore, we theoretically study the relative error of $\widehat{\mathbf{M}}_1$, which is summarized in Proposition \ref{thm2}.

\begin{algorithm}[h]
	\caption{Calculation of the Information Matrix by the Sketching Method}
	\label{alg2}
	\footnotesize
	\KwIn{The sketching number $L$ and the sketching dimension $m$\;
		\KwOut{Estimated version of $\mathbf{M}_1 :=\widetilde{\mathbf{X}}^{(T)\top}\widetilde{\mathbf{X}}^{(T)}$,$\mathbf{m}_1 := \widetilde{\mathbf{X}}^{(T)\top}\mathbf{e}^{(T)}$, $\mathbf{m}_2 := \widetilde{\mathbf{X}}^{(T)\top}\mathbf{1}_n$,
			$\mathbf{m}_3 := \overline{\widetilde{\mathbf{X}}}^{(T)\top}\mathbf{1}_n$,
			$\mathbf{M}_2 :=\overline{\widetilde{\mathbf{X}}}^{(T)\top}\overline{\widetilde{\mathbf{X}}}^{(T)}$, where $\overline{\widetilde{\mathbf{X}}}^{(T)} = \widetilde{\mathbf{X}}^{(T)} - \mathbf{1}_n\widehat{\bm{\mu}}^{(T)\top}$.}
		\For{$l =1,\cdots,L$}{
			\For{client $k =1,\cdots,K$}{
				generates its local Gaussian sketching matrix $\mathbf{S}_{k}^{A(l)}$ and $\mathbf{S}_{k}^{B(l)}$\;
				computes $\bm{\mathcal{X}}_{k}^{A(l)} = \mathbf{S}_{k}^{A(l)}\widetilde{\mathbf{X}}^{(T)}_k$, $\bm{\mathcal{X}}_{k}^{B(l)} = \mathbf{S}_{k}^{B(l)}\overline{\widetilde{\mathbf{X}}}^{(T)}_k$\;
				uploads $\bm{\mathcal{X}}_{k}^{A(l)}$ and $\bm{\mathcal{X}}_{k}^{B(l)}$ to the server\;
			}
			The server executes
			\begin{enumerate}
				\item[1.] $\bm{\mathcal{X}}^{A(l)} = (\bm{\mathcal{X}}_1^{A(l)},\ldots, \bm{\mathcal{X}}_K^{A(l)})$, $\bm{\mathcal{X}}^{B(l)} = (\bm{\mathcal{X}}_1^{B(l)},\ldots, \bm{\mathcal{X}}_K^{B(l)})$\;
				\item[2.] $\widehat{\mathbf{M}}_1^{(l)} = \bm{\mathcal{X}}^{A(l)\top}\bm{\mathcal{X}}^{A(l)}$, $\widehat{\mathbf{M}}_2^{(l)} = \bm{\mathcal{X}}^{B(l)\top}\bm{\mathcal{X}}^{B(l)}$\;
				\item[3.] generates Gaussian sketching matrix $\mathbf{S}_{0}^{(l)}$, computes
				$\widehat{\mathbf{m}}_1^{(l)} = \bm{\mathcal{X}}^{A(l)\top}\mathbf{S}_{0}^{(l)}\mathbf{e}^{(T)}$\;
				\item[4.] $\widehat{\mathbf{m}}_2^{(l)} = \bm{\mathcal{X}}^{A(l)\top}\mathbf{1}_m$\;
				\item[5.] $\widehat{\mathbf{m}}_3^{(l)} = \bm{\mathcal{X}}^{B(l)\top}\mathbf{1}_m$\;
			\end{enumerate}
		}
		The server aggregates $\widehat{\mathbf{M}}_1 = \frac{1}{L}\sum_{l=1}^{L}\widehat{\mathbf{M}}_1^{(l)}$, $\widehat{\mathbf{m}}_1 = \frac{1}{L}\sum_{l=1}^{L}\widehat{\mathbf{m}}_1^{(l)}$, $\widehat{\mathbf{m}}_2 = \frac{1}{L}\sum_{l=1}^{L}\widehat{\mathbf{m}}_2^{(l)}$,
		$\widehat{\mathbf{m}}_3 = \frac{1}{L}\sum_{l=1}^{L}\widehat{\mathbf{m}}_3^{(l)}$, $\widehat{\mathbf{M}}_2 = \frac{1}{L}\sum_{l=1}^{L}\widehat{\mathbf{M}}_2^{(l)}$\;
		Obtain $\widehat{I}_{oc}(\widehat{\bm{\theta}}^{(T)})$ by substituting $\mathbf{M}_i, \mathbf{m}_j$ ($i=1,2$ and $j=1,2,3$) in $I_{oc}(\widehat{\bm{\theta}}^{(T)})$ with their corresponding estimated version.
	}
\end{algorithm}

\setcounter{proposition}{0}
\renewcommand\theproposition{\arabic{proposition}}

\begin{proposition}\label{thm2}
	Consider $\mathbf{M}_1 = \widetilde{\mathbf{X}}^{(T)\top}\widetilde{\mathbf{X}}^{(T)}$, and $\widehat{\mathbf{M}}_1$ is an approximation of $\mathbf{M}_1$ computed by Algorithm \ref{alg2}. Then we have
	\begin{equation*}
		\frac{\|\widehat{\mathbf{M}}_1 - \mathbf{M}_1\|_2}{\|\mathbf{M}_1\|_2} = O\left( \frac{K^2n\log n}{Lm}\right).
	\end{equation*}
\end{proposition}

The detailed proof of Proposition \ref{thm2} is presented in Appendix C.4 in the supplementary materials. Proposition \ref{thm2} reveals that, the relative error of $\widehat{\mathbf{M}}_1$ depends on the client number $K$, sample size $n$, sketching number $L$, and sketching dimension $m$. The relative error would become negligible when $Lm \gg K^2n\log n$, which is also validated through our simulation study; see Appendix C.5 for details. Assume the relative error is within an acceptable level of $1/\delta$. Then based on Proposition \ref{thm2}, $Lm$ can be taken as $\delta K^2 n\log n$ (e.g., $m = K\log n, L = \delta Kn$). Denote $\widehat{I}_{oc}(\widehat{\bm{\theta}}^{(T)})$ to be the estimation of $I_{oc}(\widehat{\bm{\theta}}^{(T)})$ by using the sketching method. Then we can derive the final estimate of the asymptotic covariance as $\widehat{\mathcal{\bm{V}}} = \widehat{I}_{oc}(\widehat{\bm{\theta}}^{(T)})^{-1}\left(\mathbf{I}_d - \Gamma \right)^{-1}$. The estimated asymptotic covariance of $\widehat{\bm{\beta}}^{(T)}$ is the corresponding submatrix of $\widehat{\mathcal{\bm{V}}}$, which can be used for statistical inference without losing much statistical efficiency.

\section{Analysis of The SMEs Data}

\subsection{Data Exploration}
\label{s:visual}

\begin{figure}[h]
	\centering
	\includegraphics[width=0.6\textwidth]{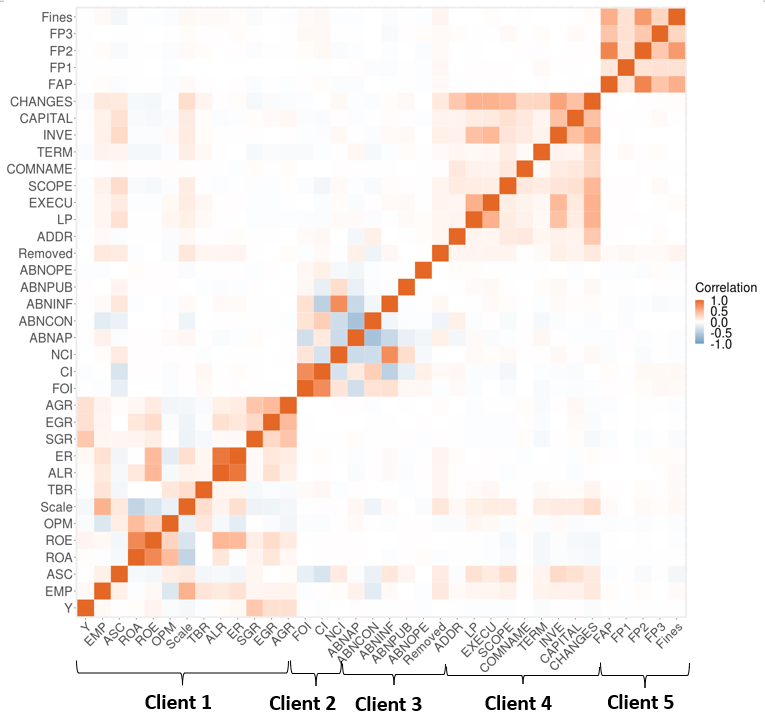} \hfill
	\caption{The rank correlation map between covariates across five clients. }
	\label{f:fy}
\end{figure}

Note that VFEM requires that variables across different clients be independent. To examine this point in the SMEs data, we first explore the correlation between variables from different clients. To this end, we compute the pairwise Spearman correlation coefficient as $r_s = \operatorname{cov}(R(X_1), R(X_2))/(\sigma_{R(X_1)}, \sigma_{R(X_2)})$, where $R(X)$ returns the rank of an arbitrary variable $X$. Figure \ref{f:fy} displays the correlation structure between covariates across different clients. As shown, covariates within the same client are more correlated than those between different clients. Specifically, the means of the absolute values of within-client correlations and between-client correlations are 0.28 and 0.03, respectively. Therefore, the assumption of independence between variables across clients is likely to be satisfied. Furthermore, the response variable NPGR has a positive correlation with covariates such as SGR (sales growth rate), EGR (growth rate of shareholder's equity), and AGR (assets growth rate).

%\newpage
\subsection{The Estimation Performance by VFEM}

We adopt the VFEM method to analyze the SMEs dataset. To implement the VFEM method, we take the estimates of CC and Impute methods as the initial values, which are denoted by CC-VFEM and Impute-VFEM, respectively. The detailed estimation results are summarized in Table \ref{table5}. The results of CC and Impute methods are also reported for easy comparison, which are the same as Table \ref{table-CI}. As shown, the VFEM method yields adjusted $R$ squares of 0.7569 (CC-VFEM) and 0.6731 (Impute-VFEM), which are substantially outperforming the CC method (0.4348) and the Impute method (0.0138). These results demonstrate better model fitting performance of VFEM over CC and Impute methods. Next, we compare the estimated coefficients of different methods. First, we find the VFEM estimator initialized with different values (i.e., CC-VFEM, Impute-VFEM) behaves nearly the same regarding the coefficient estimation. On the contrary, the CC method performs poorly with high standard deviations, which mainly owing to the large proportion of missing values. As a result, most coefficients obtained by the CC method behave insignificant. As for the Impute method, most coefficients are of different signs compared with other methods.

\begin{table}[h]
	\centering
	\caption{The estimated coefficients and R-squared for different methods on the SMEs dataset. The standard error is reported in parentheses, and * indicates the coefficient is significant under the 5\% significance level. }
	\scriptsize
	\renewcommand\arraystretch{1.2}
	\tabcolsep=0.1cm
	\label{table5}
	\begin{tabular}{@{}cccccc@{}}
		\toprule
		Client &	Variable & CC & Impute & CC-VFEM  & Impute-VFEM \\ \midrule
		\multirow{12}*{\tabincell{c}{Credit Agency}}  &	EMP & 0.0681 (0.0327)*  &  -0.0339 (0.0018)*  &	-0.0108 (0.0011)*   & -0.0105 (0.0030)* \\
		&	ASC & 0.6529 (0.6573)	&  -0.5969 (0.0204)*  &	0.6075 (0.0175)*	& 0.6203 (0.0046)*\\
		&	ROA & 0.5705 (0.3273)	&  1.4147 (0.0126)*   &	0.2552 (0.0082)*	& 0.2505 (0.0036)*\\
		&	ROE & 0.8366 (0.1932)*	&  -0.3825 (0.0083)*  &	0.1230 (0.0101)*	& 0.1717 (0.0076)*\\
		&	OPM & -0.7585 (0.6532)  &  -0.4246 (0.0198)*  &	-0.6770 (0.0184)*   &	-0.6954 (0.0122)*\\
		&	Scale & 0.0741 (0.0755)	&  -0.0755 (0.0034)*  &	0.1822 (0.0013)*    &	0.1755 (0.0009)*\\
		&	TBR & -0.0136 (0.0401)  &  0.1222 (0.0018)*   &	0.0113 (0.0042)*    &	0.0119 (0.0041)*\\
		&	ALR & -0.5904 (0.0997)* &  0.0429 (0.0047)*   &	-0.0497 (0.0084)*   &	-0.0727 (0.0040)*\\
		&	ER & 0.0675 (0.0311)*	&  0.0083 (0.0014)*   &	0.0067 (0.0053)    &	0.0079 (0.0025)*\\
		&	SGR & 0.0235 (0.0361)   &  0.0292 (0.0015)*   &	0.0086 (0.0012)*    &	0.0077 (0.0075)\\
		&	EGR & 0.0043 (0.0621)	&  0.0235 (0.0022)*   &	-0.0028 (0.0079)   &	-0.0073 (0.0017)*\\
		&	AGR & 0.1668 (0.0642)*	&  -0.0677 (0.0024)*  &	0.0155 (0.0007)*    &	0.0152 (0.0058)*\\ \midrule
		\multirow{3}*{\tabincell{c}{Market Supervision and\\ Administration Bureau}}
		&	FOI & 1.0894 (0.4355)*	&  -0.1472 (0.0940)   &	1.1478	(0.0266)*   & 1.2173 (0.0126)*\\
		&	CI & -0.9192 (0.5190)   &   0.0374 (0.0977)   &	-0.8620 (0.0305)*   &	-0.9142 (0.0403)*\\
		&	NCI & -0.9650 (0.5868)  &	0.6789 (0.1157)*  &	-0.9607 (0.0217)*   &	-0.8803 (0.0356)*\\ \midrule
		\multirow{6}*{\tabincell{c}{Judicial Organ}} &	ABNAP & -0.5648 (0.4349)   & -0.0465 (0.0553) &	-0.5021 (0.0162)*   &	-0.5250 (0.0113)* \\
		&	ABNCON & -1.5547 (0.7141)* & 0.0515 (0.1154)  &	-1.5382 (0.0225)*   &	-1.5633 (0.0352)* \\
		&	ABNINF & 1.1850 (0.6984)   & -2.6257 (0.1812)* &  1.1864 (0.0889)*  &	1.2502 (0.0628)* \\
		&	ABNPUB & -1.0497 (1.4638)  & 1.0483 (0.3778)*  & -1.0488 (0.1674)*  &	-1.0468 (0.1223)* \\
		&	ABNOPE & -2.4197 (4.2349)  & -1.3478 (0.8776)  & -2.4193 (0.3767)*  &	-2.4195 (0.2842)*\\
		&	Removed & -0.6018 (0.6071) & -0.5871 (0.1023)* & -0.5790 (0.0320)*  &	-0.5309 (0.0565)*\\ \midrule
		\multirow{9}*{\tabincell{c}{Industry and  \\Commerce Bureau}}
		&	ADDR & -0.6046 (0.7064)    & 0.9804 (0.0323)*  & -0.5262 (0.0054)*	& -0.5478 (0.0118)*\\
		&	LP & 0.4368 (0.6505)	   & 0.0323 (0.0354)   & 0.3586 (0.0078)*	& 0.3717 (0.0099)*\\
		&	EXECU & 1.1437 (0.7450)	   & -1.1334 (0.0401)* & 1.0356 (0.0128)*	& 1.0332 (0.0136)*\\
		&	SCOPE & -0.1881 (0.4515)   & -1.6775 (0.0239)* & -0.1229 (0.0045)*  &	-0.1064 (0.0098)*\\
		&	COMNAME & -0.1321 (1.2006) & 0.7315 (0.0509)*  &-0.1169 (0.0229)*	&-0.1116 (0.0105)*\\
		&	TERM & 1.4546 (0.6877)*	   & 1.6881 (0.0381)*  &1.4208 (0.0131)*	&1.4496 (0.0155)*\\
		&	INVE & 2.8771 (0.5757)*	   & -0.7099 (0.0323)* &2.6519 (0.0091)*	&2.6931 (0.0097)*\\
		&	CAPITAL & -0.1763 (0.7884) & 0.0961 (0.0361)*  &-0.2002 (0.0076)*	&-0.1994 (0.0024)*\\
		&	CHANGES & -0.2035 (0.2008) & 0.6296 (0.0096)*  &-0.4058 (0.0072)*	&-0.3896 (0.0121)*\\ \midrule
		\multirow{5}*{\tabincell{c}{Law-enforcement Agency}}
		&	FAP & 0.0829 (0.0122)	   & -0.8168 (0.0020)* &0.0264 (0.0013)*	&0.0263 (0.0141)\\
		&	FP1 & -0.0253 (1.3570)     & -0.7080 (0.1859)* & -0.0204 (0.0728)   &	-0.0222 (0.0659)\\
		&	FP2 &  -0.8555 (0.7514)    & -0.5190 (0.1158)* & -0.8139 (0.0475)*  &	-0.8371 (0.0402)*\\
		&	FP3 &   0.1968 (0.8237)	   & 2.3565 (0.1498)*  & 0.2049 (0.0331)*	&0.1956 (0.0531)*\\
		&	Fines & -0.0069 (0.5560)   & 0.0181 (0.0900)*  & 0.1402 (0.0247)*	&0.1316 (0.0255)*\\ \midrule
		$R^2$ & -- & 0.4348    & 0.0138  & 0.7569   &  0.6731  \\
		\bottomrule
	\end{tabular}
\end{table}

Since the VFEM method has a better model fitting performance, we then focus on the estimated coefficients of VFEM. Generally, the magnitudes and signs of the coefficients estimated by CC-VFEM and Impute-VFEM seem reasonable. Take the financial variables owned by the \emph{Credit Agency} as examples. The variables ROA (return on total assets) and ROE (return on equity) indicate the current profitability of SMEs likely boosts net profit growth and deserves positive coefficients.  ALR (asset liability ratio) measures debt burden; a higher value implies potential business difficulties, hence showing negative coefficients.  SGR (sales growth rate) and AGR (assets growth rate) represent enterprise development potential, which are expected to positively influence NPGR. These covariates are important financial indicators of the financial performance of SMEs, which are also verified in previous studies \citep{geng2015prediction, valaskova2018financial}.

The variables on \emph{Market Supervision and Administration Bureau} as well as \emph{Law-enforcement Agency} reflect the administrative inspection and penalties on SMEs. We observe that FOI (frequency of inspections),  FAP (frequency of administrative penalties), and  FP3 (the third category of penalty) show significant positive coefficients, revealing that intensifying supervision of SMEs could benefit their profitability.
These findings are consistent with other research. For example, \cite{wang2021does} suggested that central environmental inspection improved the total factor productivity of enterprises by promoting management efficiency and technological innovation. \cite{zhou2021does} found that environmental regulation promoted enterprise profitability through the consolidation of enterprises' cost management and elimination of small firms with high compliance costs.

The variables on \emph{Judicial Organ} mainly involve abnormality in financial and non-financial information disclosure of SMEs. As expected, the covariates ABNAP (not submit annual reports),  ABNCON (cannot be contacted at the registered address),  ABNPUB (not public company information), and ABNOPE (not engage in business activities at the registered location) all significantly negative with NPGR.
\cite{mohamad2014does} already found that non-financial information significantly influenced the firms' profitability. This is because the quality of information about firm operations and management would increase transparency and able to attract investors' confidence. On the other hand, the quality of information disclosure could influence the firms' profitability by establishing customer loyalty and developing a good corporate image.

\subsection{The Prediction Performance}

\begin{figure}[h]
	\centering
	\includegraphics[width=0.6\linewidth]{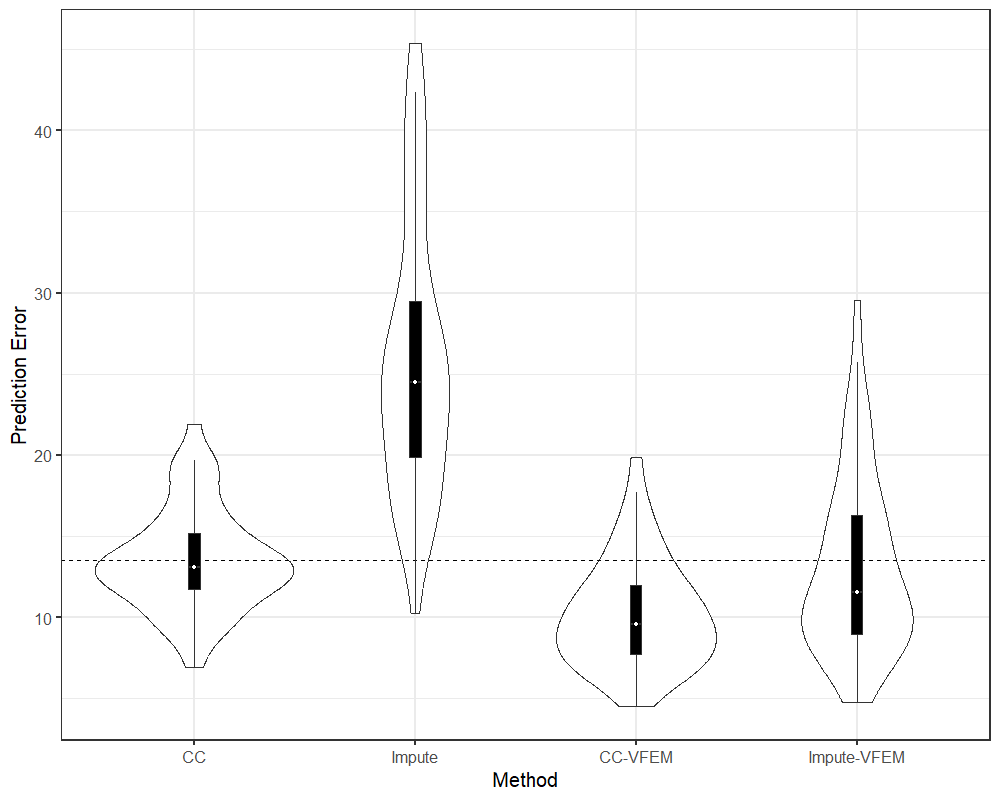}
	\caption{Prediction errors of different methods on the test dataset over $50$ random partitions.}
	\label{f2}
\end{figure}

We focus on the prediction performance of different methods in this section. To this end,  we randomly select half of the samples with complete information to form the test dataset, and the remaining samples serve as the training dataset. The procedure is repeated for $50$ times. Figure \ref{f2} reports the distribution of the prediction errors on the test dataset. It is observed that the CC method is far from satisfactory in terms of the prediction performance. The prediction error of the Impute method is even higher than the CC method.
Compared with CC and Impute, our proposed VFEM methods can greatly improve prediction accuracy by appropriately incorporating more information from the missing data.

\section{Conclusion and Discussion}

In this work, we aim to investigate the profitability of SMEs. Two main challenges are associated with this analytical task. The first one is the data isolation problem. That is, data are stored in different institutions and cannot be centralized and analyzed as a whole. Second, there exists a significant missing value issue that complicates the analysis. To address these challenges, we propose the VFEM method for the statistical estimation and inference of the linear regression model. We prove its linear convergence and establish its asymptotic normality. We further provide a computationally feasible approach to approximate the asymptotic covariance matrix to realize statistical inference. Our numerical studies demonstrate the finite sample performance of VFEM.

When applying the VFEM method to the SMEs data, we find it can improve both the estimation performance and prediction accuracy. The corresponding results also inspire several practical implications. For investors, our proposed method enables them to grasp the more accurate profitability situation of the SMEs based on more comprehensive information, adjust their investment strategies, and make prudent investments to reduce potential investment losses. Investors should especially watch out for the changes in some critical financial indicators including ROA, ALR, SGR, and AGR. For firm managers, it is imperative to establish an effective early warning system for monitoring the financial health of their enterprises. Additionally, the SMEs are advised to maintain a reasonable level of ALR, comply with laws and regulations, and improve the quality of firm information disclosure, to enhance their creditworthiness to attract more finance.

Numerous interesting directions remain for further research. Firstly, we only consider the continuous response in our study, whereas categorical response variables are often commonly encountered in practice. Hence, extending VFEM to generalized linear regression models is worth investigation. Secondly, the VFEM method is based on first-order optimization, which may be time-consuming. To accelerate the computational speed, second-order optimization methods like the Quasi-Newton algorithm are worth consideration. Thirdly, our method relies on the heavy transmission of intermediate summary statistics to facilitate localized parameter updates. This enables us to achieve lossless estimation efficiency compared with centralized analysis but at the expense of relatively heavy communication. Then how to reduce the communication costs is a good direction for future work.

\begin{funding}
Dr. Yang Li is supported by the National Natural Science Foundation of China (72271237) and Platform of Public Health \& Disease Control and Prevention, Major Innovation \& Planning Interdisciplinary Platform for the “Double-First Class” Initiative, Renmin University of China. Feifei Wang's research is supported by National Natural Science Foundation of China (No.72371241, 72171229), the MOE Project of Key Research Institute of Humanities and Social Sciences (22JJD910001), and Chinese National Statistical Science Research Project (2022LD06).

%The second author was supported in part by NIH Grant ???????????.
\end{funding}

%%%%%%%%%%%%%%%%%%%%%%%%%%%%%%%%%%%%%%%%%%%%%%
%% Supplementary Material, including data   %%
%% sets and code, should be provided in     %%
%% {supplement} environment with title      %%
%% and short description. It cannot be      %%
%% available exclusively as external link.  %%
%% All Supplementary Material must be       %%
%% available to the reader on Project       %%
%% Euclid with the published article.       %%
%%%%%%%%%%%%%%%%%%%%%%%%%%%%%%%%%%%%%%%%%%%%%%

\begin{supplement}
\stitle{Web Appendix}
\sdescription{We provide the technical proof and simulation study in Appendix.}
\end{supplement}
\begin{supplement}
\stitle{R code}
\sdescription{We provide R codes used for simulations.}
\end{supplement}

\bibliographystyle{imsart-nameyear} % Style BST file
\bibliography{reference}

\end{document}